\documentclass[10pt]{iopart}
\pdfoutput=1
\usepackage{graphicx}
\usepackage{epsf}

\begin{document}

  \vbox to 0pt{\vss
                    \hbox to 0pt{\hskip 320pt\rm LA-UR-07-1953\hss}
                   \vskip 25pt}

\title[The Cosmic Code Comparison Project]{The Cosmic Code Comparison Project}

\author{Katrin Heitmann$^{(1)}$, Zarija Luki\'c$^{(2)}$, Patricia
Fasel$^{(1)}$, Salman Habib$^{(1)}$,  Michael S. Warren$^{(1)}$,
Martin White$^{(3)}$, James Ahrens$^{(1)}$, Lee
Ankeny$^{(1)}$, Ryan Armstrong$^{(4)}$, Brian O'Shea$^{(1)}$, Paul
M. Ricker$^{(2,5)}$, Volker Springel${^{(6)}}$, Joachim Stadel$^{(7)}$,
Hy Trac$^{(8)}$}

\address{(1) Los Alamos National Laboratory, Los Alamos,
NM 87545}

\address{(2) University of Illinois, Dept. of Astronomy, Urbana, IL
61801}

\address{(3) Dept. of Astronomy, University of California Berkeley, CA
94720-3411}

\address{(4) UC Davis, Dept. of Computer
Science, Davis, CA 95616}

\address{(5) National Center for Supercomputing Applications, Urbana, IL 61801}

\address{(6) Max-Planck-Institute for Astrophysics,
85741 Garching, Germany}

\address{(7) University of Zurich,  Inst. of Theoretical Physics, 
8057 Zurich, Switzerland}

\address{(8) Princeton University, Dept. of
Astrophysical Sciences, NJ, 08544}

\ead{heitmann@lanl.gov}

\begin{abstract}
  Current and upcoming cosmological observations allow us to probe
  structures on smaller and smaller scales, entering highly nonlinear
  regimes. In order to obtain theoretical predictions in these
  regimes, large cosmological simulations have to be carried out. The
  promised high accuracy from observations make the simulation task
  very demanding: the simulations have to be at least as accurate as
  the observations. This requirement can only be fulfilled by carrying
  out an extensive code validation program. The first step of such a
  program is the comparison of different cosmology codes including
  gravitation interactions only. In this paper we extend a recently
  carried out code comparison project to include five more simulation
  codes. We restrict our analysis to a small cosmological volume which
  allows us to investigate properties of halos. For the matter power
  spectrum and the mass function, the previous results hold, with the
  codes agreeing at the 10\% level over wide dynamic ranges. We extend
  our analysis to the comparison of halo profiles and investigate the
  halo count as a function of local density. We introduce and discuss
  ParaView as a flexible analysis tool for cosmological simulations,
  the use of which immensely simplifies the code comparison task.

\end{abstract}

\maketitle

\section{Introduction}

The last three decades have seen the emergence of cosmology as
``precision science'', moving from order of magnitude estimates, to
predictions and measurements at accuracy levels better than
10\%. Cosmic microwave background observations and large galaxy
surveys have led this advance in the understanding of the origin and
evolution of the Universe. Future surveys promise even higher
accuracy, at the one percent level, over a considerably wider dynamic
range than probed earlier. In order to fully utilize the wealth of
upcoming data and to address burning questions such as the dynamical
nature of dark energy (specified by the equation of state parameter
$w=p/\rho$, $p$ being the pressure, and $\rho$ the density),
theoretical predictions must attain at least the same level of
accuracy as the observations, even higher accuracy being certainly
preferable. The highly nonlinear physics at the length scales probed,
combined with complicated gas physics and astrophysical feedback
processes at these scales, make this endeavor a daunting task.

As a first step towards achieving the final goal, a necessary
requirement is to reach the desired accuracy for gravitational
interactions alone, down to the relevant nonlinear scales.  Tests with
exact solutions such as pancake collapse~\cite{pancake} are valuable
for this task, but as shown in Ref.~\cite{Heitmann05} the results do
not easily translate into statements about the accuracy of different
simulation algorithms in realistic cosmological simulations. Exactly
solvable problems are typically highly symmetric and hence somewhat
artificial. Codes optimized for realistic situations can break down in
certain tests even if their results appear to converge in physically
relevant settings. Therefore, in order to evaluate the accuracy of
simulation codes, a broad suite of convergence and direct code
comparison tests must be carried out.

The codes used in this comparison project are all well-established,
and have been key drivers in obtaining numerous scientific
results. They are based on different algorithms and employ different
methods for error control. The code developers have already carried
out careful convergence tests themselves and verified to their
satisfaction that the codes yield reliable results. But because of the
multi-scale complexity of the dynamical problem itself, as well as the
incompleteness of most convergence tests, it is necessary to do much
more. Therefore, the aim here is to focus on comparing results from a
suite of different codes for realistic cosmological simulations. In
order to avoid uncertainties from statistical sampling, all codes are
run with exactly the same initial conditions, and all results are
analyzed using the same diagnostic tools.

The paper is organized as follows. In Section~\ref{codes} we describe
the ten simulation codes used for the comparison study. In
Section~\ref{sim} we briefly describe the simulations carried out for
this project. Next, we introduce ParaView in Section~\ref{tools}, one
of the main analysis tools used in this work. We present our results
in Section~\ref{results} and conclude in Section~\ref{conclusions}.

\section{The Codes}
\label{codes}

The ten codes used in this paper cover a variety of methods and
application arenas. The simulation methods employed include parallel
particle-in-cell (PIC) techniques (the PM codes MC$^2$ and PMM, the
PM/AMR codes Enzo and FLASH), a hybrid of PIC and direct N-body (the
AP$^3$M code Hydra), tree algorithms (the treecodes PKDGRAV and HOT),
and hybrid tree-PM algorithms ({\small GADGET-2}, TPM, and TreePM).

The PIC method models many-body evolution problems by solving the
equations of motion of a set of tracer particles which represent a
sampling of the system phase space distribution function. A
computational grid is used to increase the efficiency of the
self-consistent inter-particle force calculation. To increase dynamic
range, local force computations (e.g., P$^3$M, tree-PM) and AMR are
often used. The grid also provides a natural basis for coupling to
hydro-solvers.

Treecodes are based on the idea that the gravitational potential of a
far-away group of particles is accurately given by a low-order
multipole expansion. Particles are first arranged in a hierarchical
system of groups in a tree structure. Computing the potential at a
point turns into a descent through the tree. Treecodes naturally
embody an adaptive force resolution scheme without the overhead of a
computational grid. Tree-PM is a hybrid algorithm that combines a
long-range force computation using a grid-based technique, with
shorter-range force computation handled by a tree algorithm. In the
following we give a brief description of each code used in this
comparison study.

\subsection{The Grid Codes}

\subsubsection{MC$^2$}

The multi-species {\bf M}esh-based {\bf C}osmology {\bf C}ode MC$^2$
code suite includes a parallel PM solver for application to large
scale structure formation problems in cosmology. In part, the code
descended from parallel space-charge solvers for studying high-current
charged-particle beams developed at Los Alamos National Laboratory
under a DOE Grand Challenge~\cite{ryne98,qiang00}. MC$^2$ solves the
Vlasov-Poisson system of equations for an expanding universe using
standard mass deposition and force interpolation methods allowing for
periodic or open boundary conditions with second and fourth-order
(global) symplectic time-stepping and a Fast Fourier Transform
(FFT)-based Poisson solver. The results reported in this paper were
obtained using Cloud-In-Cell (CIC) deposition/interpolation. The
overall computational scheme has proven to be remarkably accurate and
efficient: relatively large time-steps are possible with exceptional
energy conservation being achieved.

\subsubsection{PMM}

Particle-Multi-Mesh (PMM)~\cite{Trac06} is an improved PM algorithm
that combines high mass resolution with moderate spatial resolution
while being computationally fast and memory friendly.  The current
version utilizes a two-level mesh FFT-based gravity solver where the
gravitational forces are separated into long-range and short-range
components.  The long-range force is computed on the root-level,
global mesh, much like in a PM code. To obtain higher spatial
resolution, the domain is decomposed into cubical regions and the
short-range force is computed on a refinement-level, local mesh.  This
algorithm achieves a spatial resolution of 4 times better than a
standard one-level mesh PM code at the same cost in memory.
In~\cite{Trac06}, PMM is shown to achieve very similar accuracy to
that of MC$^2$ when run with the same minimum grid spacing.

\subsubsection{Enzo}

Enzo\footnote{http://lca.ucsd.edu/codes/currentcodes/enzo} is a
publicly available, extensively tested adaptive mesh refinement (AMR),
grid-based hybrid code (hydro + N-Body) which was originally written
by Greg Bryan, and is now maintained by the Laboratory for
Computational Astrophysics at UC San Diego
~\cite{bryan97,bryan99,oshea04,oshea05}.  The code was originally
designed to do simulations of cosmological structure formation, but
has been modified to examine turbulence, galactic star formation, and
other topics of interest.  Enzo uses the Berger \& Colella method of
block-structured adaptive mesh refinement~\cite{Berger89}.  It couples
an adaptive particle-mesh method for solving the equations of dark
matter dynamics ~\cite{Efstathiou85, Hockney88} with a hydro solver
using the piecewise parabolic method (PPM), which has been modified
for cold, hypersonic astrophysical flows by the addition of a
dual-energy formalism~\cite{Woodward84, bryan95}.  In addition, the
code has physics packages for radiative cooling, a metagalactic
ultraviolet background, star formation and feedback, primordial gas
chemistry, and turbulent driving.

\subsubsection{FLASH}

FLASH~\cite{fryxell00} originated as an AMR hydrodynamics code
designed to study X-ray bursts, novae, and Type~Ia supernovae as part
of the DOE ASCI Alliances Program.  Block-structured adaptive mesh
refinement is provided via the PARAMESH library~\cite{MacNeice00}.
FLASH uses an oct-tree refinement scheme similar to~\cite{Quirk91}
and~\cite{deZP93}. Each mesh block contains the same number of zones
($16^3$ for the runs in this paper), and its neighbors must be at the
same level of refinement or one level higher or lower (mesh
consistency criterion). Adjacent refinement levels are separated by a
factor of two in spatial resolution. The refinement criterion used is
based upon logarithmic density thresholds.  Numerous extensions to
FLASH have been developed, including solvers for thermal conduction,
magnetohydrodynamics, radiative cooling, self-gravity, and particle
dynamics. In particular, FLASH now includes a multigrid solver for
self-gravity and an adaptive particle-mesh solver for particle
dynamics. Together with the PPM hydrodynamics module, these provide
the core of FLASH's cosmological simulation capabilities. FLASH uses a
variable time step leapfrog integrator. In addition to other time step
limiters, the FLASH particle module requires that particles travel no
more than a fraction of a zone during a time step.

\subsection{The Tree Codes}

\subsubsection{HOT}

This parallel tree code~\cite{ws93} has been evolving for over a
decade on many platforms. The basic algorithm may be divided into
several stages (the method of error tolerance is described
in~Ref.~\cite{sw94}).  First, particles are domain decomposed into
spatial groups. Second, a distributed tree data structure is
constructed. In the main stage of the algorithm, this tree is
traversed independently in each processor, with requests for nonlocal
data being generated as needed. A \verb-Key- is assigned to each
particle, which is based on Morton ordering. This maps the points in
3-dimensional space to a 1-dimensional list, maintaining as much
spatial locality as possible.  The domain decomposition is obtained by
splitting this list into $N_p$ (number of processors) pieces. An
efficient mechanism for latency-hiding in the tree traversal phase of
the algorithm is critical. To avoid stalls during nonlocal data
access, effectively explicit `context switching' is done using a
software queue to keep track of which computations have been put aside
waiting for messages to arrive. This code architecture allows HOT to
perform efficiently on parallel machines with fairly high
communication latencies~\cite{warren03}. HOT has a global time
stepping scheme. The code was among the ones used for the original
Santa Barbara Cluster Comparison Project~\cite{frenk99} and also
supports gas dynamics simulations via a smoothed particle
hydrodynamics (SPH) module~\cite{fw02}.

\subsubsection{PKDGRAV}

The central data structure in PKDGRAV~\cite{stadel} is a tree
structure which forms the hierarchical representation of the mass
distribution. Unlike the more traditional oct-tree which is used in
the Barnes-Hut algorithm~\cite{barnes} and is implemented in HOT,
PKDGRAV uses a k-D tree, which is a binary tree. The root-cell of this
tree represents the entire simulation volume. Other cells represent
rectangular sub-volumes that contain the mass, center-of-mass, and
moments up to hexadecapole order of their enclosed regions. PKDGRAV
calculates the gravitational accelerations using the well known
tree-walking procedure of the Barnes-Hut algorithm. Periodic boundary
conditions are implemented via the Ewald summation
technique~\cite{ewald}. PKDGRAV uses adaptive time stepping.  It runs
efficiently on very large parallel computers and has produced some of
the world's highest resolution simulations of cosmic structures. A
hydrodynamics extension called GASOLINE exists.

\subsection{The Hybrid Codes}

\subsubsection{Hydra}

HYDRA~\cite{couchman95} is an adaptive P$^3$M (AP$^3$M) code with
additional SPH capability.  In this paper we use HYDRA only in the
collisionless mode by switching off gas dynamics. The P$^3$M
method combines mesh force calculations with direct summation of
inter-particle forces on scales of two to three grid spacings. In
regions of strong clustering, the direct force calculations can become
significantly expensive. In AP$^3$M, this problem is tackled by
utilizing multiple levels of subgrids in these high density regions,
with direct force computations carried out on two to three spacings of
the higher-resolution meshes. Two different boundary conditions are
implemented in HYDRA, periodic and isolated.  The time step algorithm
in the dark matter-only mode is equivalent to a leapfrog algorithm.

\subsubsection{\small GADGET-2}

The N-body/SPH code {\small GADGET-2} \cite{Springel2001,Springel2005}
employs a tree method \cite{barnes}, to calculate gravitational
forces. Optionally, the code uses a tree-PM algorithm based on an
explicit split in Fourier space between long-range and short-range
forces \cite{Bagla2002}. This combination provides high performance
while still retaining the full spatial adaptivity of the tree
algorithm, allowing the code to reach high spatial resolution
throughout a large volume. By default, {\small GADGET-2} expands the
tree multipoles only to monopole order, in favor of a compact tree
storage, a cache-optimized tree-walk, and consistent and efficient
dynamic tree updates. The cell-opening criterion used in the tree walk
is based on an estimator for the relative force error introduced by a
given particle-cell interaction, such that the tree force is accurate
up to a prescribed maximum relative force error. The latter can be
lowered arbitrarily, if desired, at the expense of higher calculation
times.  The PM part of {\small GADGET-2} solves Poisson's equation on
a mesh with standard fast Fourier transforms, based on a CIC mass
assignment and a four-point finite differencing scheme to compute the
gravitational forces from the potential. The smoothing effects of grid
assignment and interpolation are corrected by an appropriate
deconvolution in Fourier space. The time-stepping of {\small GADGET-2}
uses a leap-frog integrator which is symplectic in case constant
timesteps (in the log of the expansion factor) are employed for all
particles. However, the code is normally run in a mode where
individual and adaptive timesteps are used to speed up the calculation
time. To this end, the timesteps for the short-range dynamics are
allowed to freely adapt to any power of two subdivision of the
long-range timestep. {\small GADGET-2} is fully parallelized for
massively parallel computers with distributed memory, based on the MPI
standard.  The code can also be used to simulate hydrodynamical
processes using the particle-based smoothed particles hydrodynamics
(SPH) method (e.g.~\cite{Monaghan1992}), in an entropy conserving
formulation \cite{Springel2002}, a feature which is however not
exercised in the simulations considered in this paper.

\subsubsection{TPM}

TPM~\cite{xu95,bode00} is a publicly-available hybrid code combining a
PM and a tree algorithm. The density field is broken down into many
isolated high-density regions using a density threshold
criterion. These contain most of the mass in the simulation but only a
small fraction of the volume. In these regions, the gravitational
forces are computed with the tree algorithm while for the bulk of the
volume the forces are calculated via a PM algorithm, the PM time steps
being large compared to the time-steps for the tree-algorithm. The PM
algorithm uses the CIC deposition/interpolation scheme and solves the
Poisson equation using FFTs.The time integrator in TPM is a standard
leap-frog scheme: the PM time steps are fixed whereas tree particles
have individual time steps, half of the PM step or smaller.

\subsubsection{TreePM}

The algorithmic structure of the TreePM code~\cite{white02} is very
similar to {\small GADGET-2}.  The particles are integrated using a
second-order leap-frog method, with position and canonical momentum as
the variables.  The time step is dynamically chosen as a small
fraction (depending on the smoothing length) of the local free-fall
time and particles have individual time steps.  The force on any given
particle is computed in two stages.  The long-range component of the
force is computed using the PM method, while the short range component
is computed from a global tree.  A spline softened force law is used.
The tree expands forces to monopole order only, and cells are opened
based upon the more conservative of a geometric and relative force
error criterion. The PM force is computed by direct FFT of the density
grid obtained from CIC mass assignment.

\section{The Simulations}
\label{sim}

A previous code comparison suite~\cite{Heitmann05} considered three
cosmological test problems: the Santa Barbara Cluster~\cite{frenk99},
and two large-scale structure simulations of $\Lambda$CDM models in a
64$h^{-1}$Mpc box and a 256$h^{-1}$Mpc box. In the latter two cases,
the primary target of this previous work was to investigate results in
a medium resolution regime, addressing statistical quantities such as
the two-point correlation function, the density fluctuation power
spectrum, and the dark matter halo mass function.

In this paper we focus further attention on one of these tests, the
smaller of the $\Lambda$CDM boxes. Due to the small box size, the
force resolution of all codes -- including the pure mesh codes -- is
in principle sufficient to analyze properties of individual halos
themselves. This allows us to extend the dynamic range of the code
comparison to higher resolution than studied earlier. In this new
regime, we expect to see a much broader divergence of results because
of the more demanding nature of the test. (Even in the previous
analysis~\cite{Heitmann05}, the power spectrum was unexpectedly
deviant at the larger wavenumbers considered.) Our aim is to
characterize the discrepancies and attempt to understand the
underlying causes.

All codes were given exactly the same particle initial conditions at a
redshift $z_{\rm in}=50$. The initial linear power spectrum was
generated using a fit to the transfer function~\cite{kh97}, a
modification of the BBKS fit~\cite{bbks86}. This fit does not capture
baryon oscillations but takes baryonic suppression into account (these
details are of only limited relevance for the test). The cosmology
underlying the simulations is given by $\Omega_{\rm CDM}=0.27$,
$\Omega_{\rm b}=0.044$, $\Omega_{\Lambda}=0.686$, $h=0.71$,
$\sigma_8=0.84$, and $n=0.99$. The simulation was run with 256$^3$
particles, which leads to an individual particle mass of
$m_p$=1.362$\cdot 10^9h^{-1}$M$_\odot$.

While performing a comprehensive code comparison study which involves
very different algorithms -- such as grid and particle-based methods
in the present case -- a central and difficult question immediately
arises: what is the most informative way to compare the codes and
learn from the results? The difficulty is compounded by the fact that
codes are often optimized under different criteria and controlling
numerical error is a complex multi-parameter problem in any case, even
for codes that share the same general underlying algorithm.

As a case in point, let us consider the choice of force resolution for
each code. (Since the volume and number of particles are fixed, the
mass resolution is the same for each run.) One option would be to run
all codes with the same formal force resolution but this, aside from
wasting resolution for the high-resolution codes, also suffers from
the problem that it is not easy to compare resolutions across
different algorithms; moreover, time-stepping errors also must be
folded into these sorts of estimates. Finally, such a comparison would
be rather uninteresting, because realistic cosmological simulations
are run with higher resolutions than would be possible in a
conservative test of this type: Interesting effects on small scales
would be missed. A more uncontrolled, but nevertheless useful option
is to allow every simulator to run her or his code with close to the
optimal settings they would also use for a scientific run (given the
other restrictions imposed by the test problem). In this case, a more
realistic comparison can be performed in which we can access the
robustness of conclusions from cosmological simulations. Here, while
our approach adheres more closely to the second strategy, we do try to
assess at what length scales one should expect a specific code to
break down assuming that the resolution of the code is accurately
estimated by the simulator.

\begin{center}
\begin{table*}
\caption{\label{tab1} Softening lengths measured in
$h^{-1}$kpc. The different smoothing kernels have been converted into
Plummer softening equivalents by matching the potential at the
origin. While this procedure is only approximate, it makes a
comparison of the different force resolutions more meaningful. For
details on the conversion see the main text.}
\footnotesize\rm
\begin{tabular}{cccccccccc}
\br
MC$^2$ & PMM    & Enzo   & FLASH & HOT &  PKDGRAV & Hydra & {\small GADGET-2} &
TPM & TreePM   \\
\mr
62.5 & 62.5 & 62.5  & 62.5 & 7.1  &
1.6     & 28.4 & 7.1     &  5.1 & 5.7 \\ 
\br
\end{tabular}
\end{table*}
\end{center}

The nominal resolutions for the different codes for the performed runs
are as given in Table~\ref{tab1}. We have converted the different
softening kernels into Plummer equivalents following the normalization
conventions of Ref.~\cite{dehnen}. We have matched the different
softening kernels $\phi$ at zero and compared them at this point. With
the normalization conventions in Ref.~\cite{dehnen}, we find:
\begin{eqnarray}
\phi_{\rm Plummer}(0)&\propto& \frac{1}{\epsilon},\\
\phi_{\rm Spline}(0)&\propto& \frac{7}{5}\frac{1}{\epsilon},\\
\phi_{\rm K_3}(0)&\propto& \frac{2079}{512}\frac{1}{\epsilon},
\end{eqnarray}
where $\epsilon$ is the softening length. The grid resolution of the
PM and AMR codes is roughly equivalent to the Plummer softening. HOT
and Hydra have Plummer force kernels implemented, PKDGRAV uses
Dehnen's K$_3$ kernel~\cite{dehnen} and the three tree-pm codes use
spline kernels. With the above definitions, it is easy to convert the
spline and K$_3$ kernels into Plummer via
\begin{eqnarray}
\epsilon_{\rm Spline}&=&1.4\epsilon_{\rm Plummer},\\
\epsilon_{\rm K_3}&=&4.06\epsilon_{\rm Plummer},
\end{eqnarray}
which we used to standardize the force resolution quotes in
Table~\ref{tab1}. We note that some of the codes below could have been
run at higher resolution, and the values below should not be thought
of as resolution limits. In fact, the choices of these values
represent compromises due to run time considerations as well as a
(loosely) pre-planned scatter to try and determine the effects of
force resolution on the simulation results.

\section{Analysis Framework and Tools}
\label{tools}

Broadly speaking, the major aim of our code comparison project is to
characterize differences in results from large cosmological simulation
codes, identify the causes underlying these differences, and, if
possible, develop strategies to reduce or eliminate the differences in
order to obtain reliable results over large length and mass scales.
If it is not possible to eliminate some of the differences, e.g. due
to insufficient force resolution in grid codes, it is still important
to provide robust criteria that correctly determine the scales at
which the code can be trusted, and with what accuracy.

The identification and characterization of differences in code results
is not as straightforward as it first appears. Certainly we may, and
do, compare standard statistical quantities such as low-order
correlation functions. However, there is much more information in the
data beyond this, e.g., large-scale structure morphology, substructure
in the density field, and subtle features such as the variation in
halo bias as a function of the local environment. For these reasons,
it is often very useful to simply {\em look} at the region or object
of interest in the simulation and compare it across the different
codes. Following this qualitative comparison -- perhaps even inspired
by it -- the aim is to construct a hypothesis about the cause for the
perceived difference which then has to be carefully tested with
quantitative measures.

It is very desirable to have a framework which combines these two
steps in a convenient, and eventually seamless, manner. The framework
should allow differences and anomalies picked up by eye from data sets
to be immediately queried and quantified using a programmable
toolkit. An example relevant to cosmological simulations is the
following. Suppose we ``see'' fewer halos in one simulation compared
to another. The analysis tool should then provide quantitative
information of the following type: in what areas is the difference
larger, what is the exact number count of the halos in this region,
what is the difference in the environment (e.g. by comparing the local
density in the two codes), what is the halo history in the region, and
so on.

As part of this paper we include an introduction of
ParaView~\cite{paraview}~\footnote{We use ParaView 2.6 throughout this
  paper. This is the latest stable release which can be downloaded at
  http://www.paraview.org/HTML/Download.html} to the cosmology and the
wider computational communities. ParaView has some of the features
discussed above built in and allows the user to implement additional
analysis tools.  ParaView is an open-source, scalable visualization
tool which is designed as a layered architecture. The foundation and
first layer of ParaView is the visualization toolkit (VTK). VTK
provides data representations, algorithms, and a mechanism to
interconnect these to form a working program. The second layer is a
parallel extension to the visualization toolkit which supports
streaming of all data types and parallel execution on shared and
distributed memory machines. The third layer is ParaView
itself. ParaView provides a graphical user interface and transparently
supports the visualization and rendering of large datasets via
hardware acceleration, parallelism, and level-of-detail techniques.

For the code comparison project we have implemented a particle reader
which works with the data format used throughout this paper. This
allows other simulators who wish to test their codes against our
results to use exactly the same analysis tool. As explained later, we
have also implemented a diverse set of diagnostic tools relevant for
cosmological simulations. These help to ease the analysis of large
simulation data sets and make it more efficient. We plan to extend the
set of available analysis features in the near future.

\section{Results}
\label{results}
\subsection{Results for the Full Simulation Box}

\begin{figure}
\center\includegraphics[width=80mm,angle=-90]{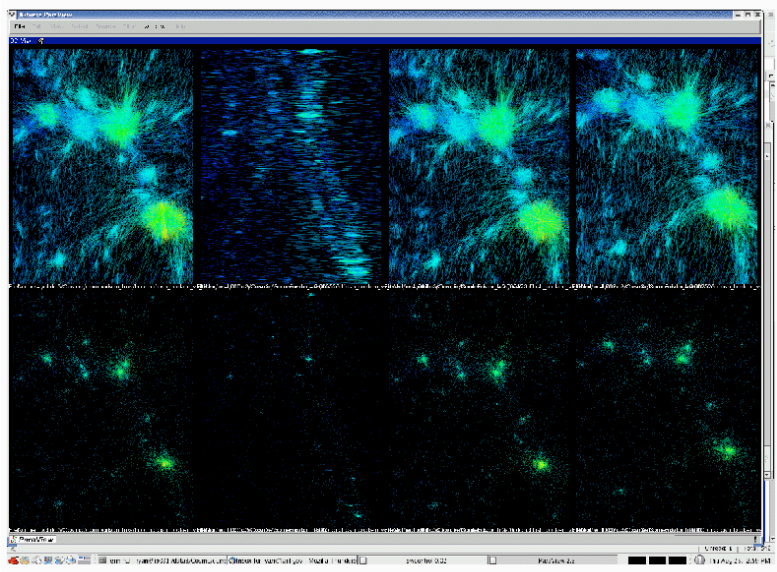}
\caption{\label{snap}Screenshot of the comparative visualization manager in
ParaView. Upper row: results from four different codes, zoomed into
a dense region of the simulations. Particles are displayed as arrow
glyphs, colored with respect to their velocity magnitude. Lower row:
same region, the particles now displayed simply as dots. }
\end{figure}

As an initial test, a simple view of the simulation output at $z=0$
proves to be very useful. ParaView offers a comparative visualization
option in which the results from different simulations can be shown
simultaneously.  Manipulation on any one output in this mode results
in the same manipulation for all the others. ParaView allows fly-ins,
rotation of the box, projections, and has many more features which
make it convenient to inspect the outcome of the simulation.  A
screenshot of the comparative visualization manager is displayed in
Figure~\ref{snap} -- a zoom into an arbitrary region of the simulation
box showing simultaneous results from four different codes. In the
upper row a subset of the particles is shown as arrow glyphs, colored
by velocity magnitude, the lower row shows the particles as dots with
the same coloring scheme. A quick inspection of these snapshots
reveals that the code 2 run had a problem with the velocities and code
4 had slightly incorrect boundary conditions (the whole picture being
shifted upward). (Of course these initial bugs were fixed before going
on to the final results discussed below!)

\begin{figure*}[t]
\begin{center}
\parbox{14cm}{
\parbox[t]{14cm}{
\parbox[t]{6.5cm}
{\begin{center}
\hspace{0cm}\includegraphics[width=65mm]{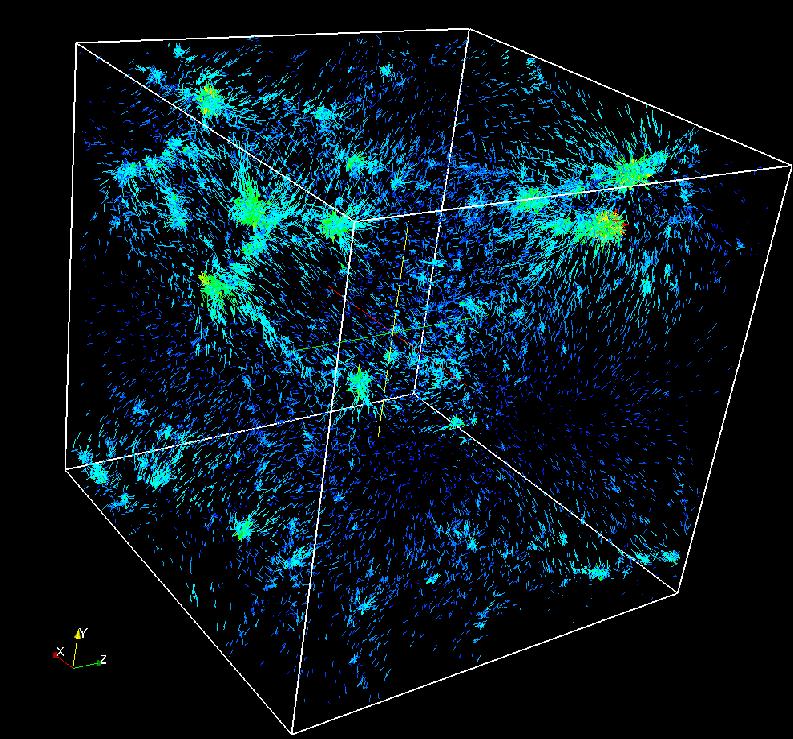}
\end{center}}
\parbox[t]{6.5cm}
{\begin{center}
\hspace{0cm}\includegraphics[width=65mm]{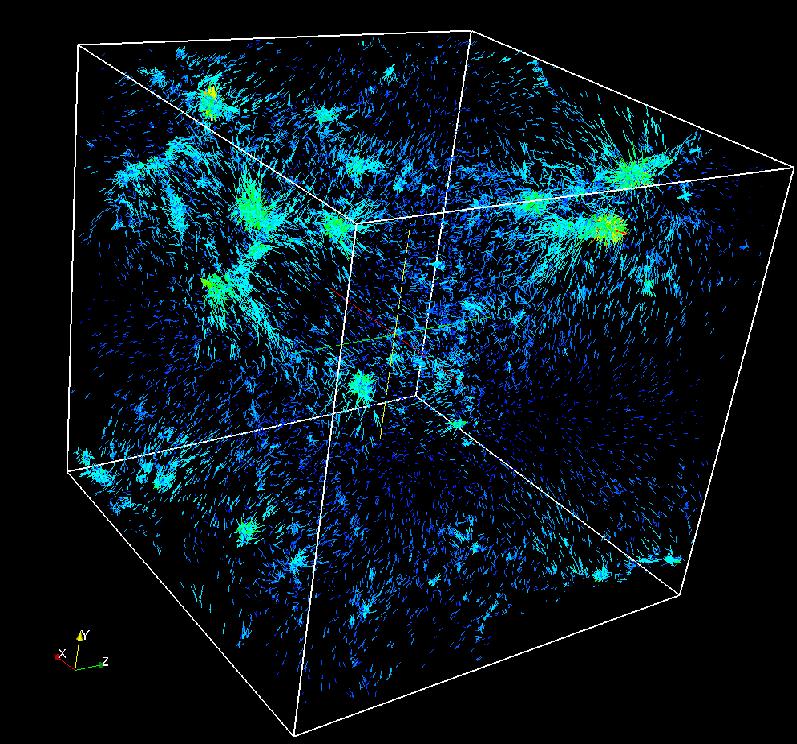}
\end{center}}}}
\end{center}
\caption{A subset of the 20,000 particles at $z=0$ from the {\footnotesize
GADGET-2} simulation (left) and the Enzo simulation (right). The
particles are shown with vector arrow glyphs which are sized and
colored by their velocity magnitude (blue: slowest, red: fastest).}
\label{fig:box}
\end{figure*}

Figure~\ref{fig:box} shows a comparison of the final {\small GADGET-2}
and Enzo outputs. We show a subsample of 20,000 particles, each
displayed with vector arrow glyphs, sized and colored by their
velocity magnitude.  The arrow glyphs nicely represent the flows in
the box to the major mass concentrations. As to be expected, particles
in the field are slow (blue), while the particles in the halos have
the largest velocities (yellow to red).  While the overall appearance
of both simulations shown is very similar, subtle differences can be
seen (e.g., there are no small structures in the flow regions in the
Enzo simulations), indicating the higher resolution employed in the
{\small GADGET-2} run. (Five of the biggest halos in the simulation
will be examined in more detail below, the resolution differences
becoming significantly more apparent.)

\subsection{Dark Matter Halos}

The halo paradigm is central to any large-scale structure analysis;
dark matter in simulations, discretized in the form of heavy
collisionless particles, forms clearly visible filaments (stripes) and
halos (clumps of dark matter) through the process of gravitational
instability. Figure~\ref{fig:box} shows these structures clearly for
the simulations studied in this paper. This picture agrees well with
observations of galaxy rotation curves, and velocity dispersions of
galaxies in clusters which favor scenarios where luminous, baryonic
matter is embedded in massive, extended, and close to spherical
conglomerates of dark matter. In simulations, dark matter halos can be
identified and ``weighed'' in different ways. We can measure
overdensities (see e.g.~\cite{halof}) or use group finding algorithms
such as friend-of-friend (FOF) algorithms~\cite{fof} to find halos.

A remarkable feature of halos was found by Navarro, Frenk, and
White~\cite{NFW}: dark matter halos of all masses, from dwarf galaxies
to the largest clusters of galaxies, have spherically-averaged density
profiles that are well-described by a single, ``universal'' formula
\begin{equation}
\rho(r) = \frac{\rho_c \delta_c}{r/r_s 
             \left(1+r/r_s\right)^2}.
\label{eqn:nfw}
\end{equation}
The scaling radius, $r_s$, and characteristic overdensity, $\delta_c$,
are free parameters of the model, while $\rho_c$ is the critical
density for closure of the Universe.  Even though a theoretical
explanation for this universal profile has not been found, the nature
of the profile itself has received extensive support from simulations
carried out by many different groups~\cite{nfwconf} (there do remain
questions about the behavior at very small radii, but these are not
relevant here).

Here we are interested in the variation of the profiles produced by
the different codes, tending towards the outer region of the
halo. This variation may be significant for determining halo masses
via the often used FOF algorithm. The mass that the halo finder will
``see'', strongly depends on the density and density gradient close to
the virial radius ($R_{200}$) of a halo.  On the other hand,
accurately reproducing the inner slope of a halo profile is the prime
test of the code's force resolution. On scales below this resolution
limit, particle positions get randomized, resulting in a flattened
density profile (numerical errors can also lead to a sharpening of the
profile due to an associated unphysical damping).

\begin{figure*}
\begin{center}
\parbox{14cm}{
\parbox[t]{14cm}{
\parbox[t]{7cm}
{\begin{center}
\hspace{0cm}\includegraphics[width=90mm,angle=0]{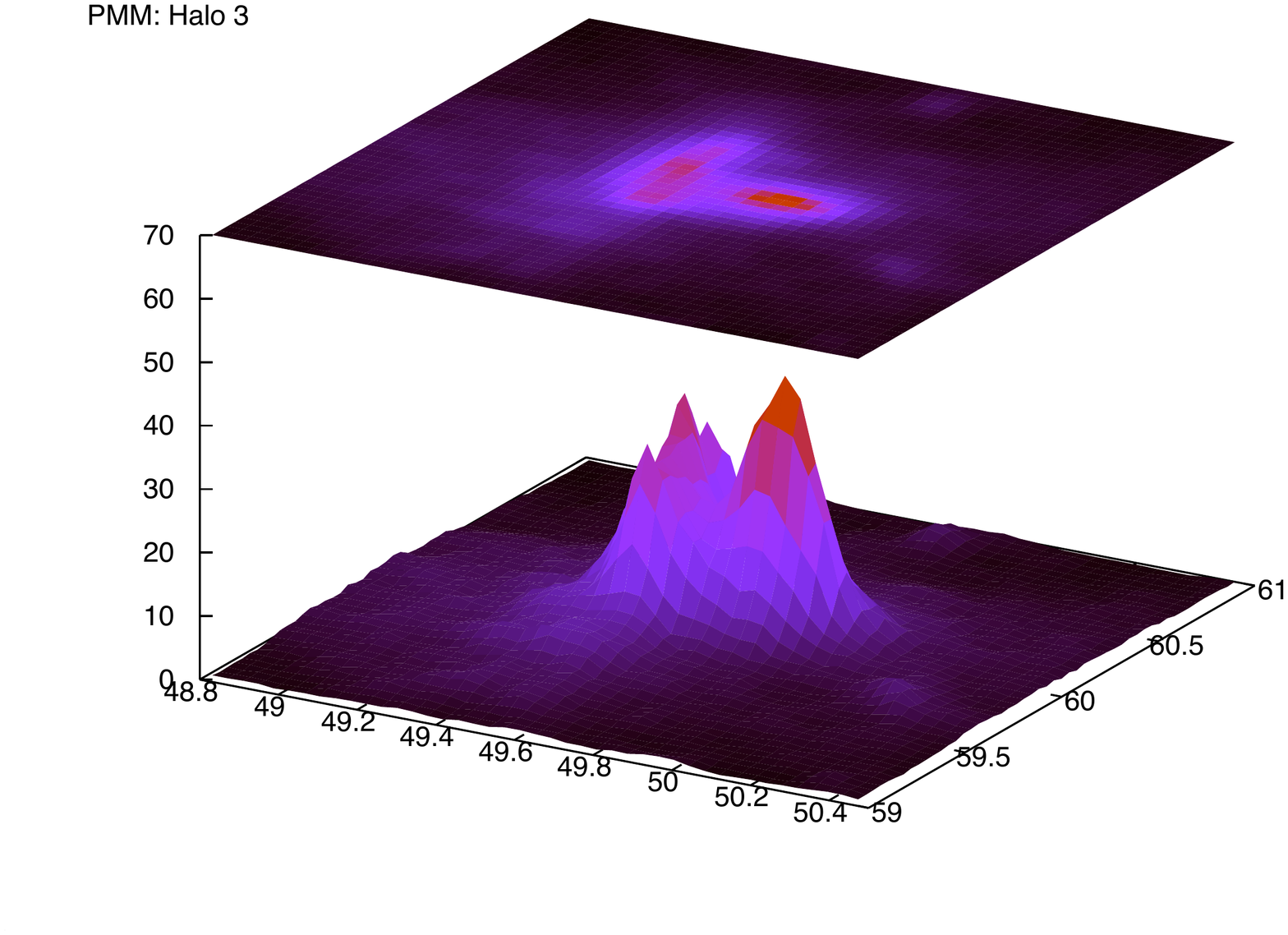}
\end{center}}
\parbox[t]{7cm}
{\begin{center}
\hspace{0cm}\includegraphics[width=90mm,angle=0]{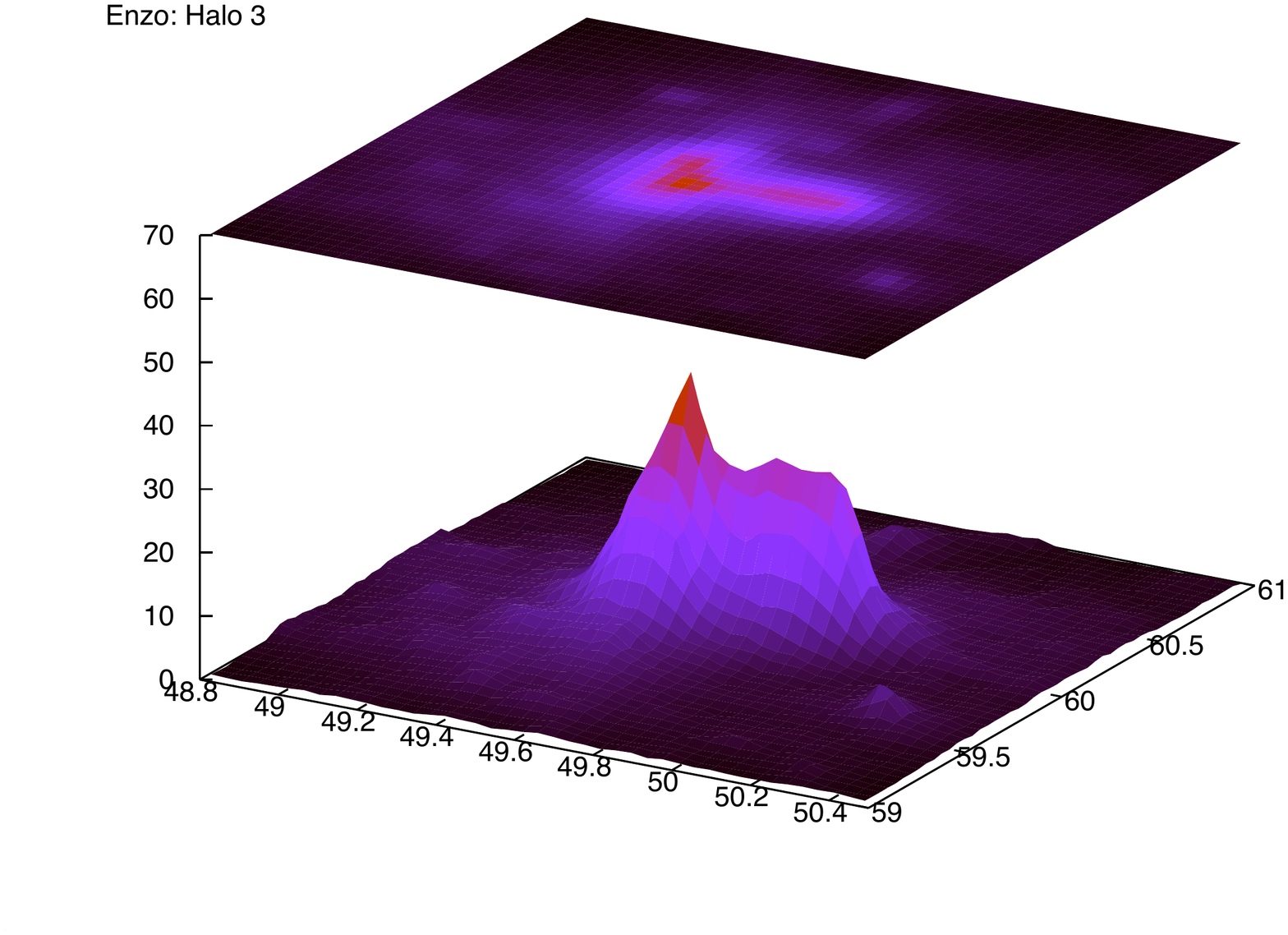}
\end{center}}}

\vspace{-1cm}

\parbox[t]{14cm}{
\parbox[t]{7cm}
{\begin{center}
\hspace{0cm}\includegraphics[width=90mm,angle=0]{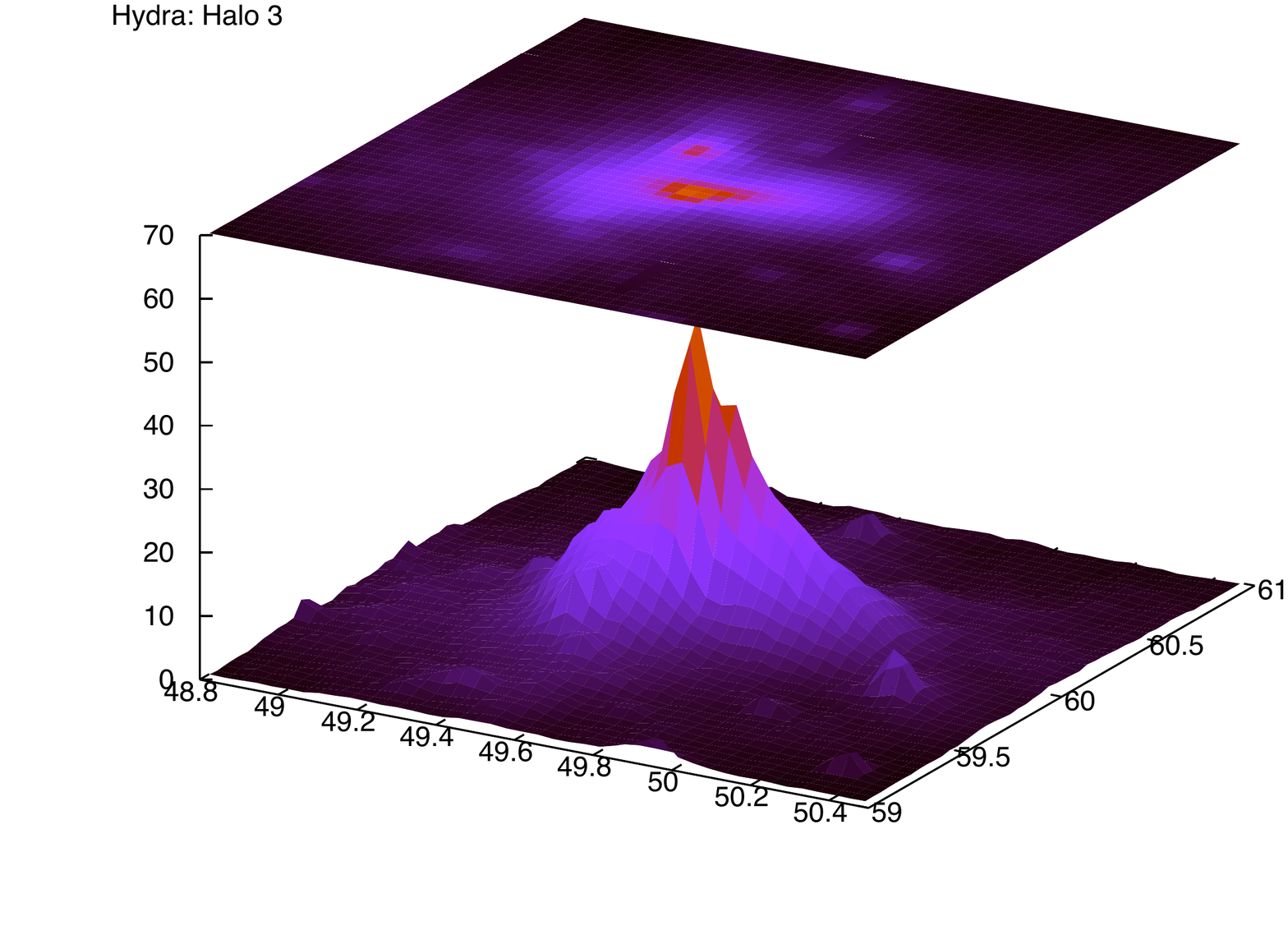}
\end{center}}
\parbox[t]{7cm}
{\begin{center}
\hspace{0cm}\includegraphics[width=90mm,angle=0]{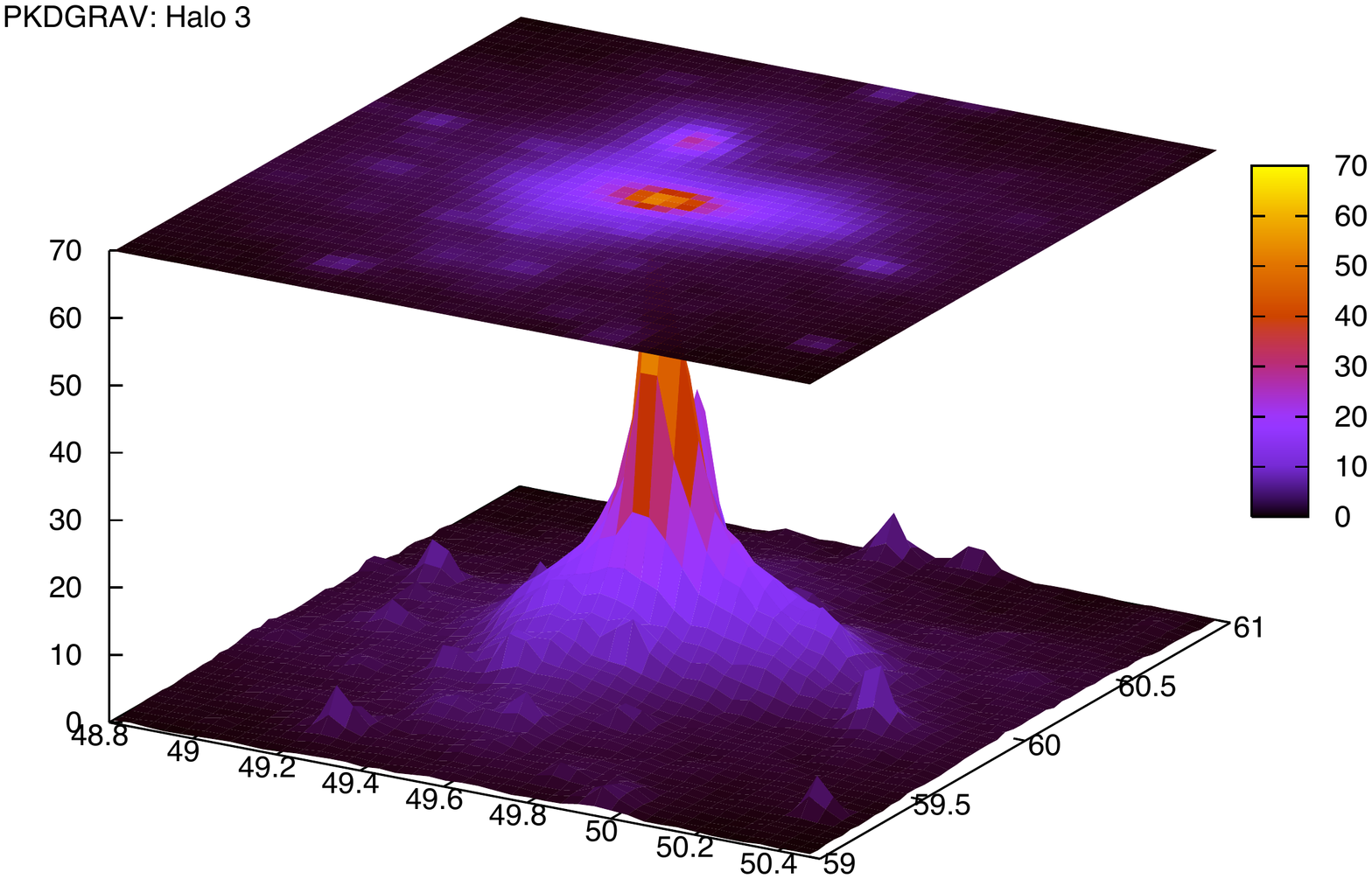}
\end{center}}}

\vspace{-1cm}

\parbox[t]{14cm}{
\parbox[t]{7cm}
{\begin{center}
\hspace{0cm}\includegraphics[width=90mm,angle=0]{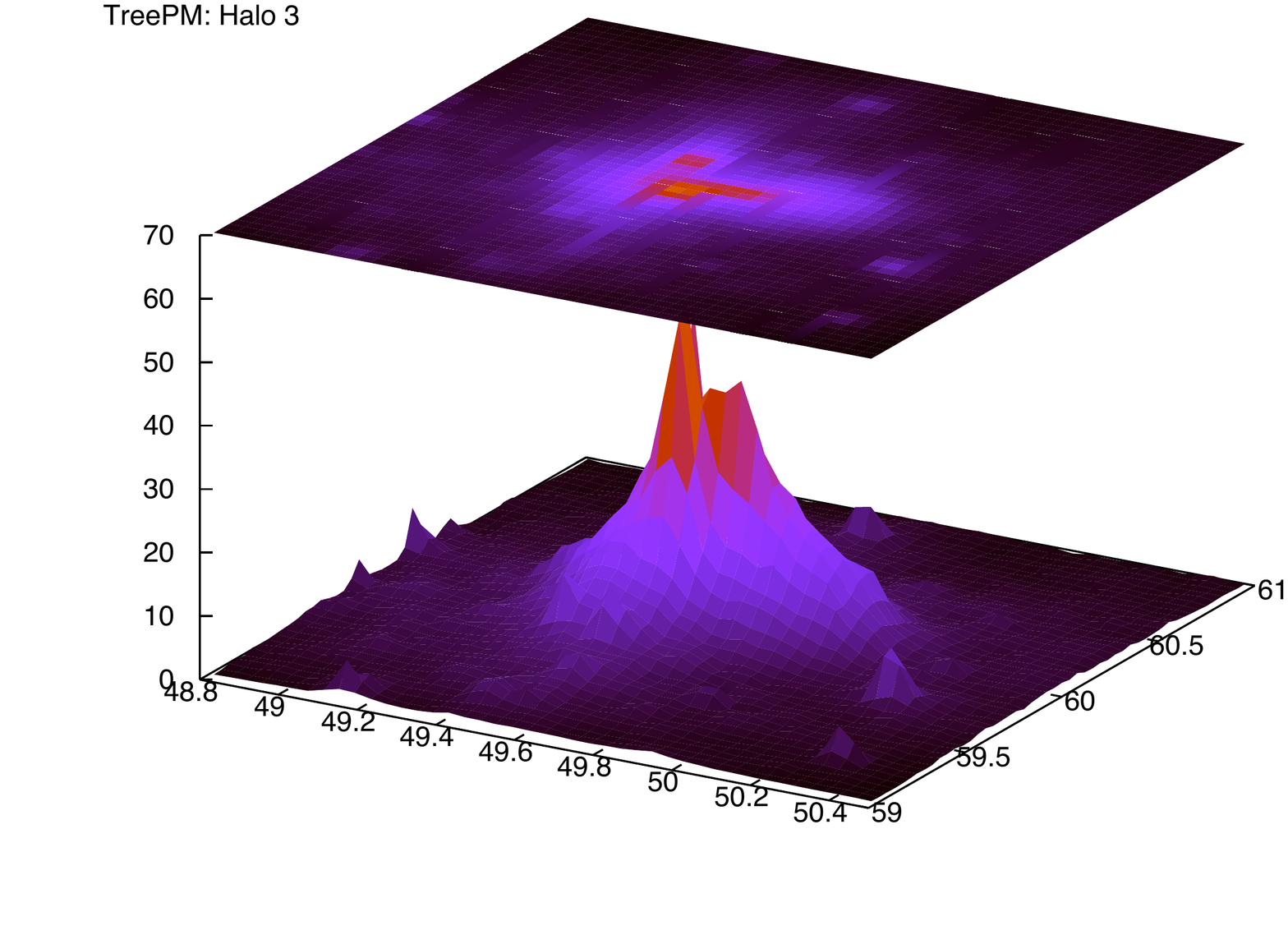}
\end{center}}
\parbox[t]{6.5cm}
{\begin{center}
\hspace{0cm}\includegraphics[width=90mm,angle=0]{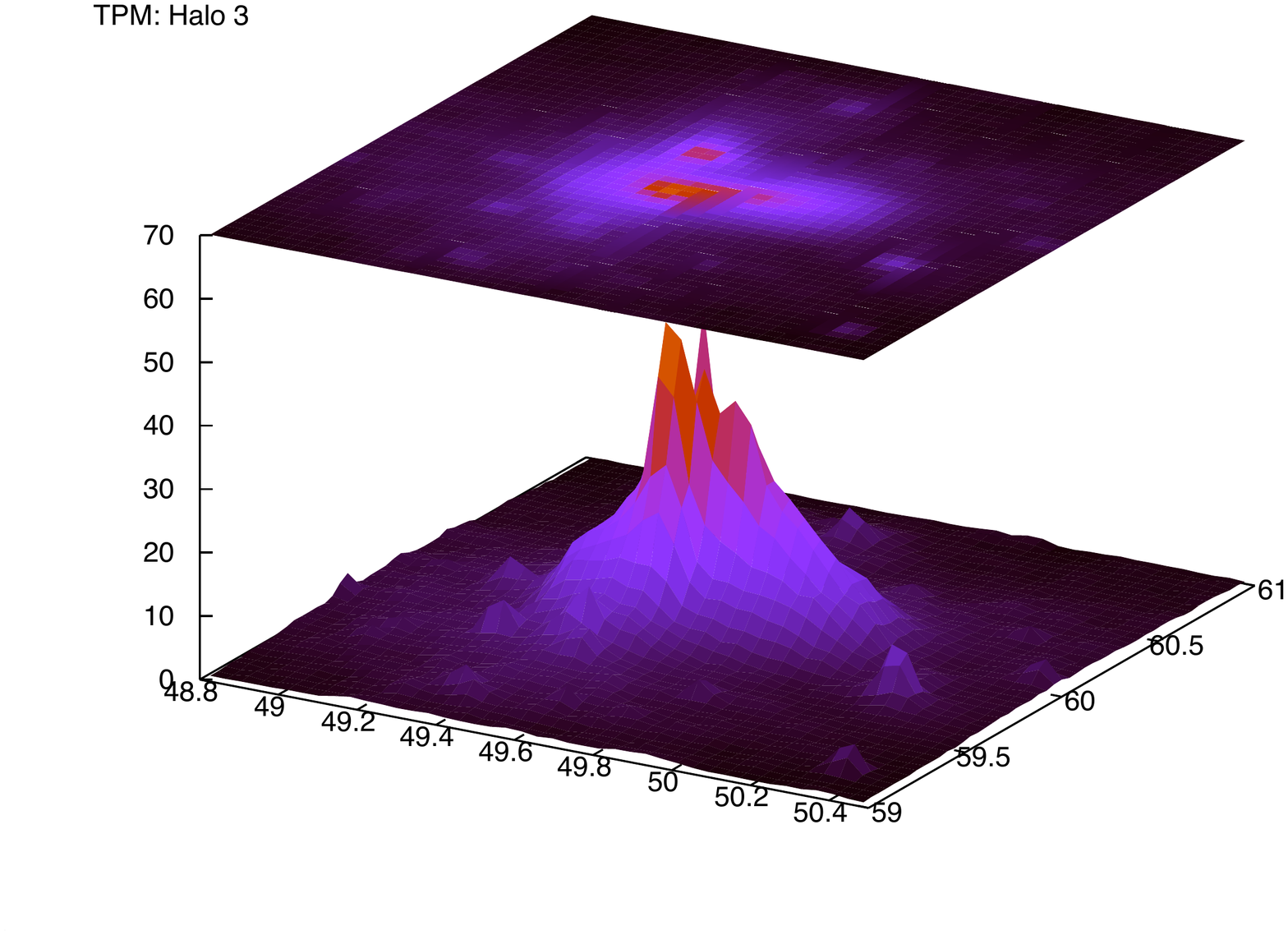}
\end{center}}}
}
\end{center}
\caption{\label{halo3d}Two-dimensional densities from PMM, Enzo, Hydra,
PKDGRAV, TreePM, and TPM for Halo 3. The panel on the top of each
graph shows the projected density. The color coding is the same for
each plot, shown in the result for PKDGRAV.}
\end{figure*}

We first compare the five heaviest halos from the simulations; their
masses range between approximately 2 to $5\cdot 10^{14} h^{-1}
M_{\odot}$, thus each halo is sampled with 150,000 or more
particles. The individual halo masses (as found by the FOF algorithm)
are in agreement within 3\% for all ten codes. Note that the FOF
masses found for the grid codes are slightly higher. This is
presumably due to their lower resolution in this comparison, resulting
in less tight halos. The FOF halo finder can identify more particles
in the fuzzier outskirts of lower resolution simulations as belonging
to the halo than in the high resolution runs. The centers of the halos
are defined by the minimum of the local potential of the halo. Here
the agreement among the codes is even better than for the masses --
the difference is less than 0.5\% of the box size. In Table~\ref{tab2}
we show the center and mass of one of the halos, Halo 3. This halo
(also shown in Figure~\ref{halo3d}) has the size and mass of a group
of galaxies. The dispersion in the mass and position of the center is
similar for the other halos, whose profiles we investigate next.

\begin{center}
\begin{table*}
\centering
\caption{\label{tab2} Halo 3 data: distance of the center from the mean value 
for all codes, and the mass of the halo from different simulations.}
\footnotesize\rm
\begin{tabular}{c | c c c c}
\br
Code & $\Delta X_c$ [kpc/h] & $\Delta Y_c$ [kpc/h] & $\Delta Z_c$ [kpc/h] & Mass [$10^{14}$ M$_{\odot}$/h]  \\
\mr
MC$^2$ & -86.23 & 158.81 & -14.68 & 2.749 \\ 
\mr
PMM & 201.68 & 33.90 & 10.24 &  2.757 \\
\mr
Enzo & -21.36 & 45.16 & 11.36 & 2.745 \\
\mr
FLASH & -41.66 & -22.56 & -23.10 & 2.726 \\
\mr
HOT & -30.02 & -120.54 & 43.99 & 2.720 \\
\mr
PKDGRAV & 38.58 & 52.19 & -43.98 & 2.679 \\
\mr
Hydra & 19.91 & -28.29 & 0.77 & 2.721 \\
\mr
{\small GADGET-2} & -27.08 & -59.00 & -0.70 & 2.705 \\
\mr
TPM & -36.37 & -35.09 & 1.04 & 2.697 \\
\mr
TreePM & -17.45 & -24.62 & 13.63 &  2.727 \\
\br
\end{tabular}
\end{table*}
\end{center}

\begin{figure*}
\begin{center}
\parbox{14cm}{
\parbox{14cm}{
\parbox{7.cm}
{\begin{center}
\hspace{0cm}\includegraphics[width=70mm]{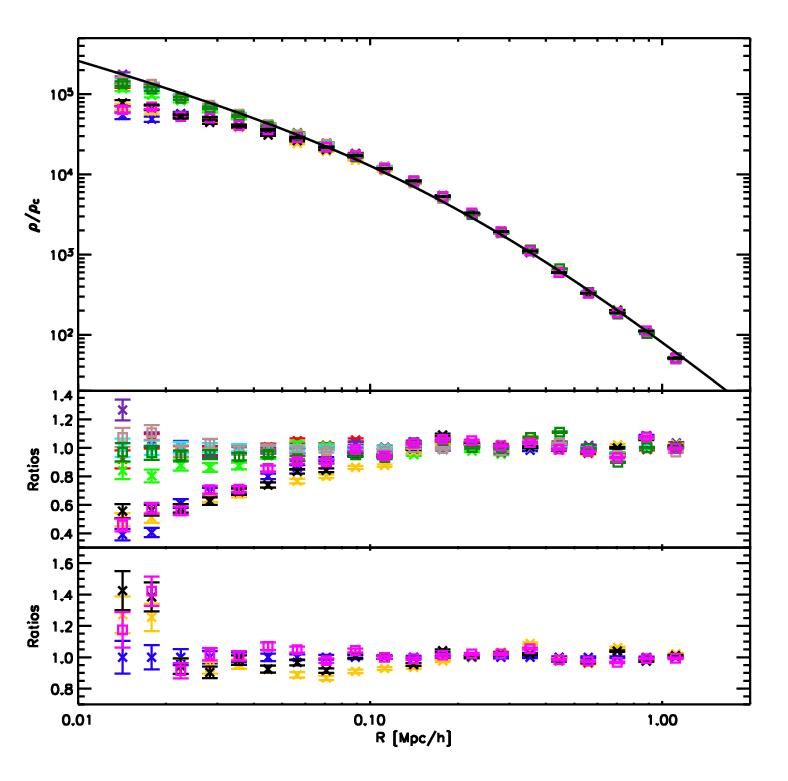}
\end{center}}
\parbox{7.cm}
{\begin{center}
\hspace{0cm}\includegraphics[width=70mm]{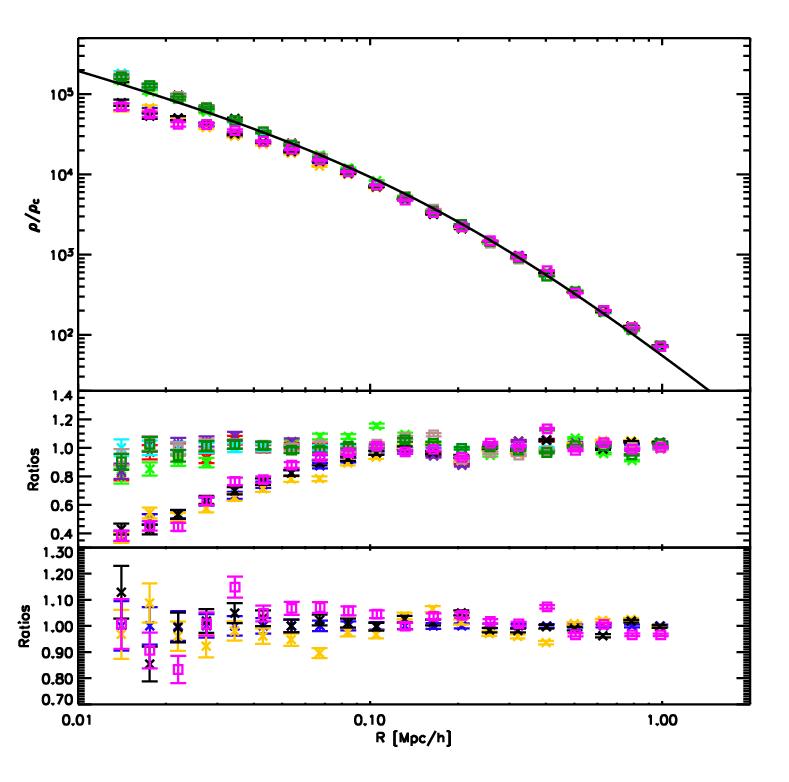}
\end{center}}}

\vspace{-.8cm}

\parbox{14cm}{
\parbox{7cm}
{\begin{center}
\hspace{0cm}\includegraphics[width=70mm]{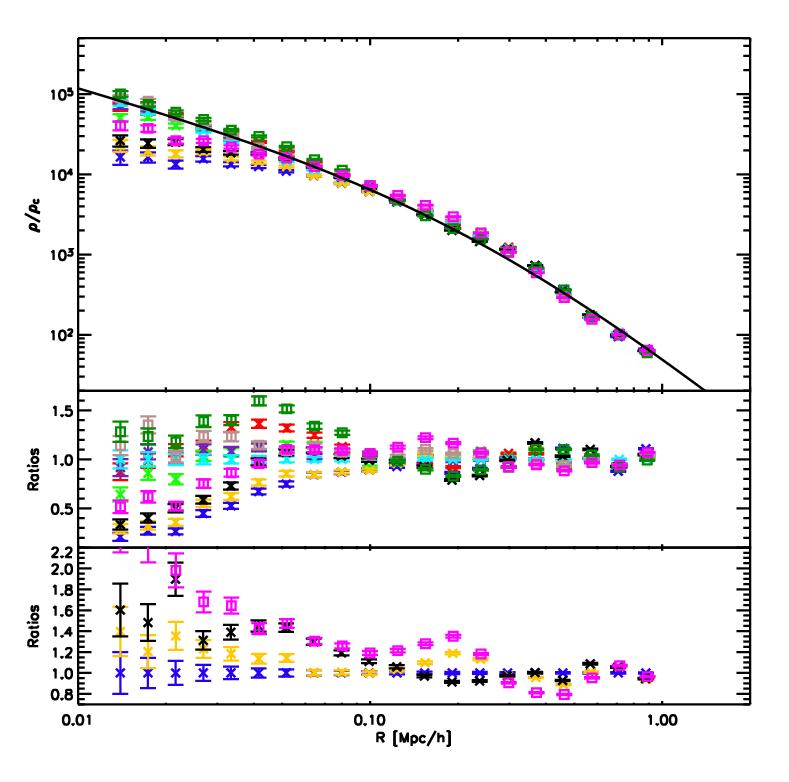}
\end{center}}
\parbox{7cm}
{\begin{center}
\hspace{0cm}\includegraphics[width=70mm]{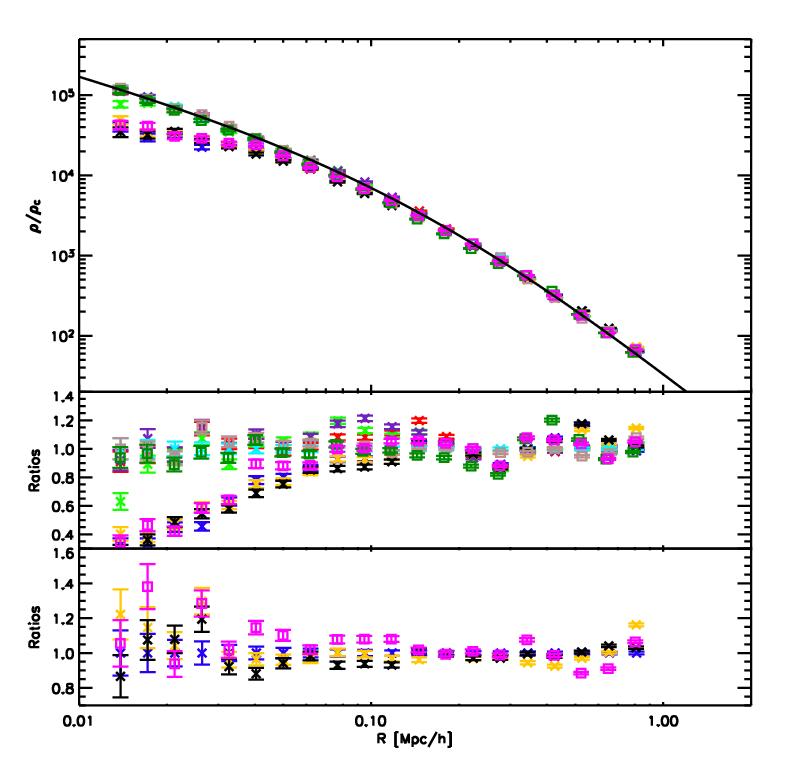}
\end{center}}}

\vspace{-0.8cm}

\parbox{14cm}{
\parbox{7cm}
{\begin{center}
\hspace{0cm}\includegraphics[width=70mm]{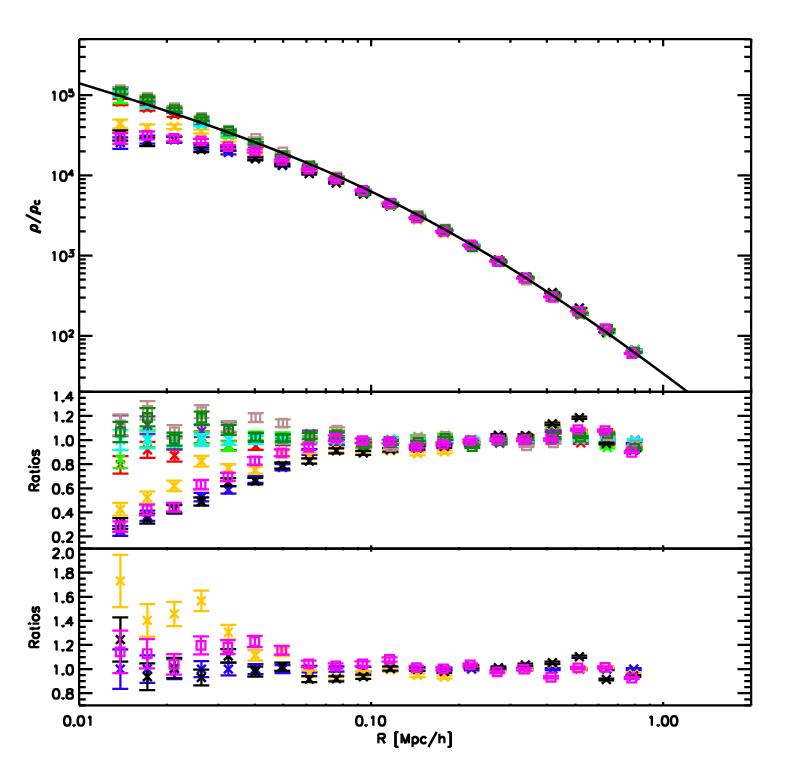}
\end{center}}
\parbox{7cm}
{\begin{center}
\hspace{0cm}\includegraphics[width=70mm]{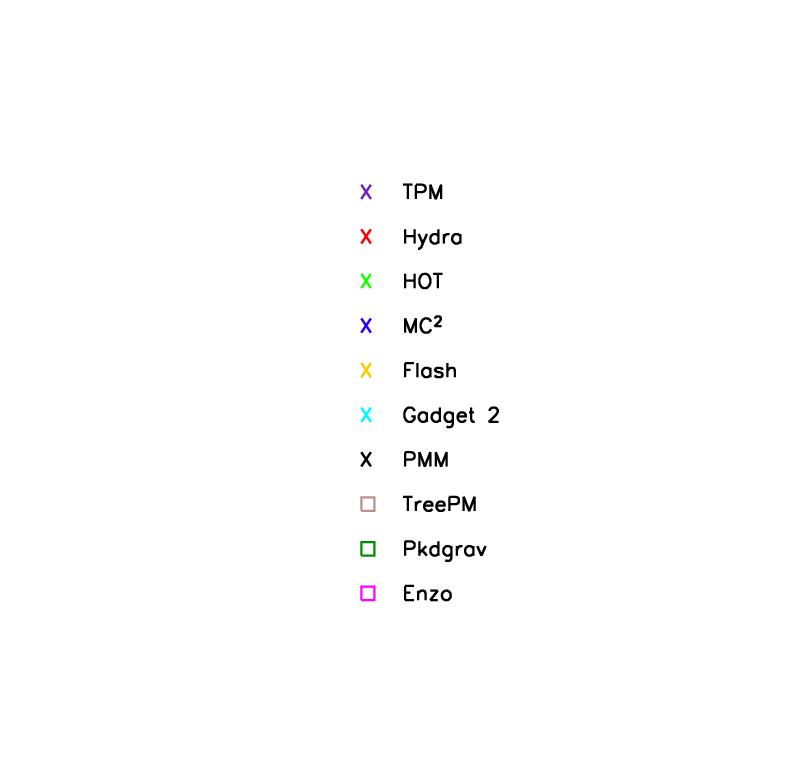}
\end{center}}}
\caption{\label{profiles}Halo profiles for the five heaviest halos in
  the simulation. The black line shows the best-fit NFW profile to the
  TPM simulation, mainly to guide the eye. In the outer regions all
  codes agree very well. In the inner regions the fall-off of the grid
  codes is as expected due to resolution limitations. The fall-off
  point can be predicted from the finite force resolution and agrees
  well with the results. The middle panel in each plot shows the ratio
  of the different codes with respect to {\footnotesize GADGET-2}. The
  lower panels show only the four grid codes and the ratio with
  respect to MC$^2$.}}
\end{center}
\end{figure*}

In Figure~\ref{profiles} we present the spherically averaged density
profiles for the five heaviest halos in the simulation. As an
arbitrary reference, the black line represents the best NFW fit
(Equation~\ref{eqn:nfw}) for the TPM data. The fit is shown up to the
inner 10 $h^{-1}$kpc of each halo. In addition, we show two residual
panels for each halo profile. The upper panel shows the ratio of all
codes with respect to {\small GADGET-2}, while the lower panel shows
only the four grid codes and ratios with respect to MC$^2$.

The agreement in the outer part of the halos is excellent. As
expected, the codes exhibit different behaviors on small scales
(depending on their force resolution and time-stepping), thus the
inner parts of halos are not always the same. While the high
resolution codes successfully track the profile all the way in to the
plotting limits of Figure~\ref{profiles}, the profiles from the mesh
codes depart much earlier (60-100 $h^{-1}$kpc), with approximately
constant density in the core.  The onset of the flattening is
consistent with the nominal resolution of the grid codes, which is
given in Table.~\ref{tab1}.  Note that among the mesh codes there is
no significant difference between the fixed mesh codes which ran at
the highest resolution throughout the whole simulation volume, and the
AMR codes whose base mesh spacing is a factor of 4 times lower.

\begin{figure*}
\begin{center}
\parbox{12cm}{
\parbox[t]{12cm}{
\parbox[t]{6.cm}
{\begin{center}
\hspace{0cm}\includegraphics[width=60mm,angle=0]{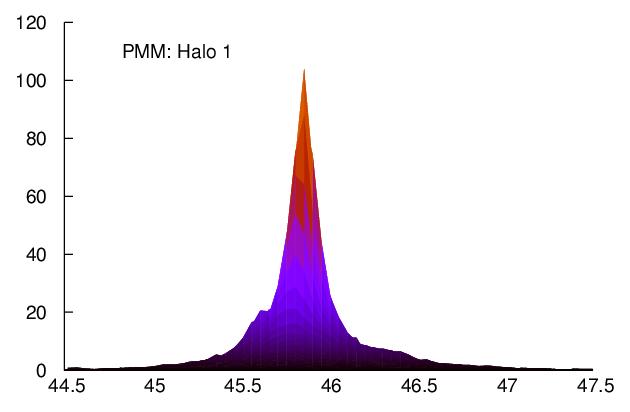}
\end{center}}
\parbox[t]{6.cm}
{\begin{center}
\hspace{0cm}\includegraphics[width=60mm,angle=0]{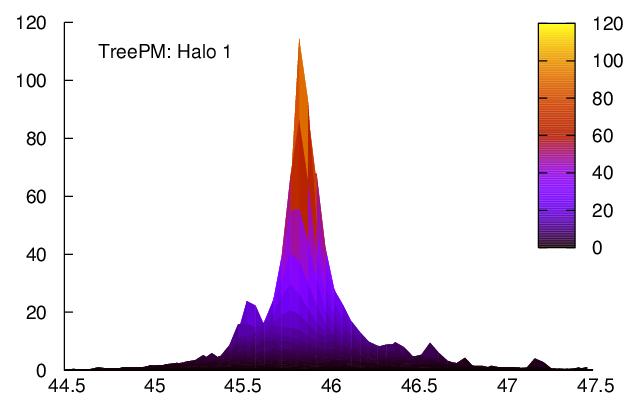}
\end{center}}}
\caption{\label{halo1}Projected and normalized two-dimensional density for 
Halo 1 from PMM (left) and TreePM (right). TreePM has a slightly
higher density in the inner region of the halo than PMM, as to be
expected from the different force resolutions. Overall the agreement
is very good.}}
\end{center}
\end{figure*}

We now study three of the five halos in more detail, restricting
attention to particles within a sphere of radius $2\cdot R_{200}$. The
profiles of the largest halo, Halo 1, shown in Figure~\ref{profiles},
agree well down to $R=0.06 h^{-1}$Mpc; at smaller scales the finite
resolution of the grid codes becomes apparent. Nevertheless, the grid
codes and the high-resolution codes among themselves yield very
consistent results. Figure~\ref{halo1} shows the density of Halo 1 for
the lower resolution code PMM and the higher resolution code TreePM in
two-dimensional projection. The two-dimensional density field is
computed on a 100$\times$100 grid within the $2\cdot R_{200}$ region,
projected onto the $z$-direction (another projection along the
$x$-direction is also shown). The projected density field has been
normalized by dividing out the mean density in this area. The mean
density is very close across the different codes, hence the
normalization allows for direct comparisons of the projected density
fields.  As mentioned earlier, the positions of the halo centers
(density peaks) are in remarkably good agreement. Due to its higher
resolution, the density in the center of the halo from the TreePM run
is slightly higher (as to be expected from the profiles). In addition,
TreePM shows slightly more substructure on the outskirts of the halo,
displayed by the small ``hills''. Overall, the halo is very smooth and
well defined, which is reflected in the good agreement of the
profiles. The density plots for the four grid codes are very
similar. The small structures around the halo in the other codes also
show only very minor variations, thus the PMM and TreePM results can
be considered to be representative.

\begin{figure*}
\begin{center}
\parbox{13cm}{
\parbox{13cm}{
\parbox{6.5cm}
{\begin{center}
\hspace{0cm}\includegraphics[width=60mm,height=60mm,angle=0]{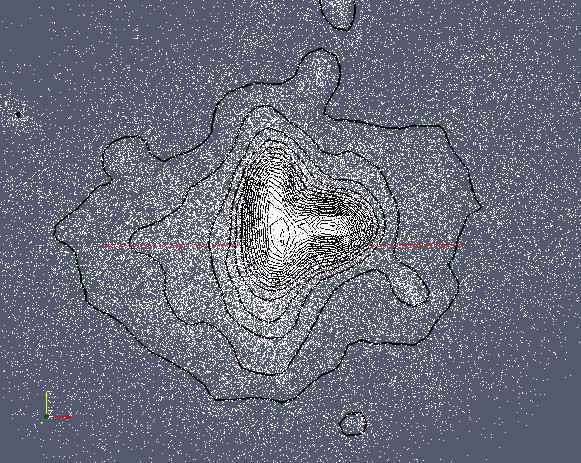}
\end{center}}
\parbox{6.5cm}
{\begin{center}
\hspace{0cm}\includegraphics[width=60mm,height=60mm,angle=0]{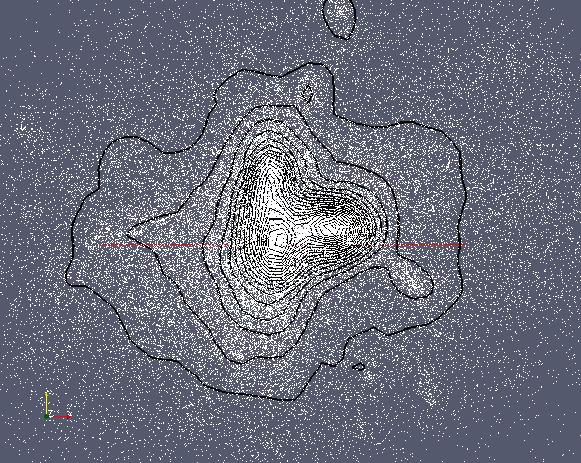}
\end{center}}}
\parbox{13cm}{
\parbox{6.5cm}
{\begin{center}
\hspace{0cm}\includegraphics[width=60mm,height=60mm,angle=0]{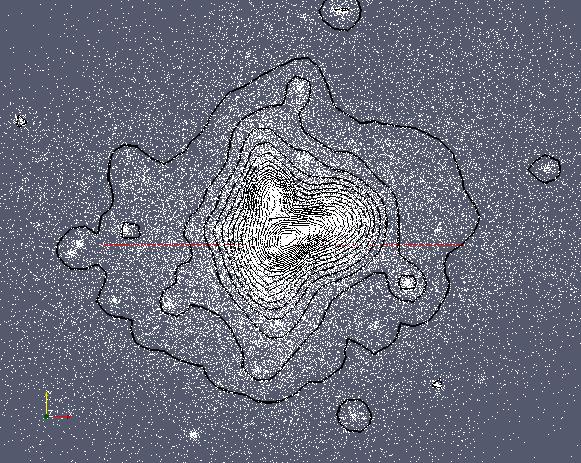}
\end{center}}
\parbox{6.5cm}
{\begin{center}
\hspace{0cm}\includegraphics[width=60mm,height=60mm,angle=0]{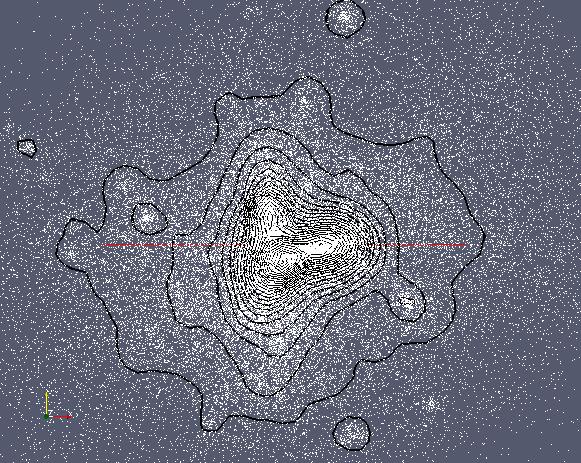}
\end{center}}}
}
\caption{\label{halo3b}Two-dimensional contour plot of the projected
  density for Halo 3 from MC$^2$, FLASH, {\footnotesize GADGET-2}, and
  HOT (left upper to right lower plot). White: particles, black:
  contour smoothed with a Gaussian Filter. }
\end{center}
\end{figure*}

\begin{figure*}
\begin{center}
\parbox{15cm}{
\parbox{15cm}{
\parbox{3.5cm}
{\begin{center}
\hspace{0cm}\includegraphics[width=35mm,angle=0]{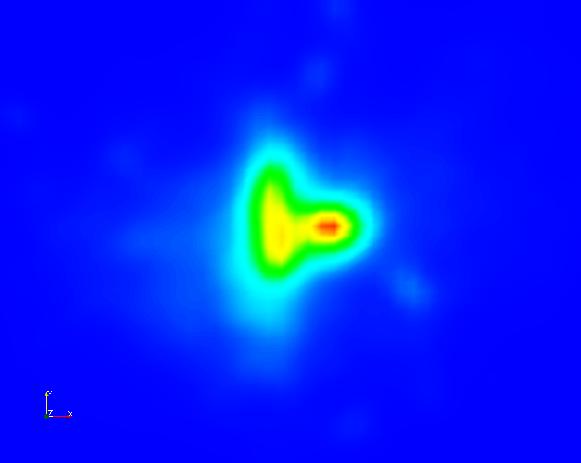}
\end{center}}
\parbox{3.5cm}
{\begin{center}
\hspace{0cm}\includegraphics[width=35mm,angle=0]{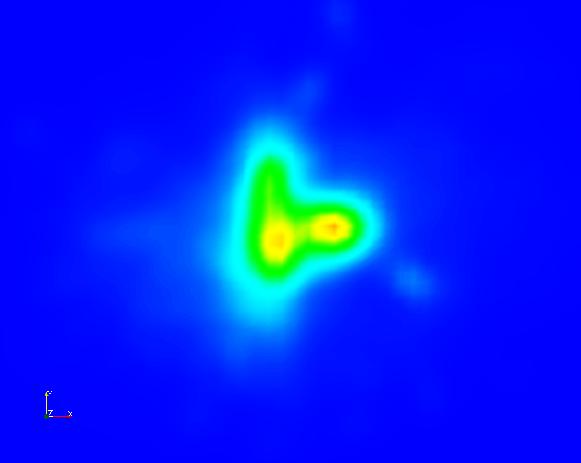}
\end{center}}
\parbox{3.5cm}
{\begin{center}
\hspace{0cm}\includegraphics[width=35mm,angle=0]{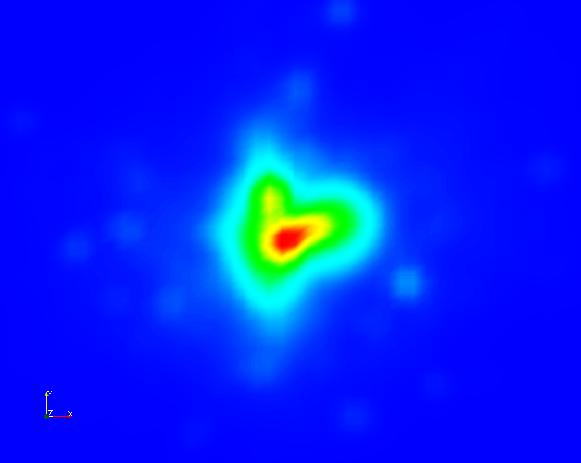}
\end{center}}
\parbox{3.5cm}
{\begin{center}
\hspace{0cm}\includegraphics[width=35mm,angle=0]{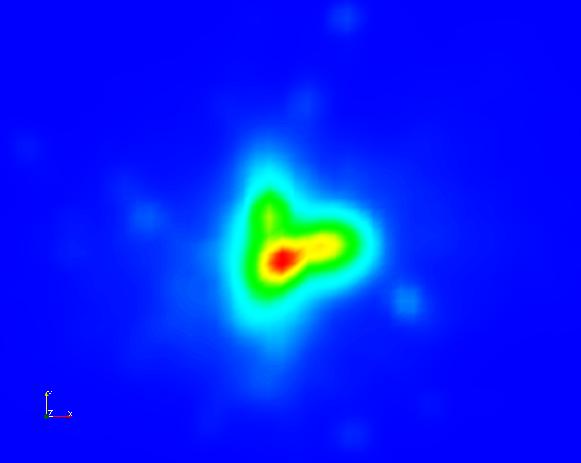}
\end{center}}}
}
\caption{\label{halo3splash}Same as in Figure~\ref{halo3b}: MC$^2$,
FLASH, {\footnotesize GADGET-2}, and HOT.}
\end{center}
\end{figure*}

The profiles of Halo 3 show substantially more variation among the
different codes in the inner region, relative to the other four
halos. Studying it in more detail, we first investigate a subset of
four codes: MC$^2$, FLASH, {\small GADGET-2}, and HOT, covering a wide
range of force resolutions. In Figure~\ref{halo3b} we show a zoom into
the center of the halo. The particles are shown in white. Superimposed
on the particle distribution is a 2-dimensional density contour
evaluated on a 100$\times$100 grid and smoothed with a Gaussian
filter, projected along the $z$-direction. (The contouring and
filtering are intrinsic functions in ParaView.\footnote{ParaView
provides a Gaussian filter called vtkGaussianSplatter. This is a
filter that injects input points into a structured points (volume)
dataset. As each point is injected, it ``splats'' or distributes values
to nearby voxels. Data is distributed using an elliptical, Gaussian
distribution function. The distribution function is modified using
scalar values (expands distribution) or normals (creates ellipsoidal
distribution rather than spherical).  In general, the Gaussian
distribution function $f(x)$ around a given splat point $p$ is given
by $f(x) = {\rm ScaleFactor} \cdot \exp( {\rm ExponentFactor}((r/{\rm
Radius})^2) )$ where $x$ is the current voxel sample point; $r$ is
the distance $|x-p|$, ExponentFactor $\le$ 0.0, and ScaleFactor can be
multiplied by the scalar value of the point $p$ that is currently
being splatted. If points normals are present (and NormalWarping is
on), then the splat function becomes elliptical (as compared to the
spherical one described by the previous equation). The Gaussian
distribution function then becomes: $f(x) = {\rm ScaleFactor} * \exp(
{\rm ExponentFactor}*( ((rxy/E)^2 + z^2)/R^2) )$ where $E$ is a
user-defined eccentricity factor that controls the elliptical shape of
the splat; $z$ is the distance of the current voxel sample point along
normal $N$; and $rxy$ is the distance of $x$ in the direction
perpendicular to $N$.  This class is typically used to convert
point-valued distributions into a volume representation. The volume is
then usually iso-surfaced or volume rendered to generate a
visualization. It can be used to create surfaces from point
distributions, or to create structure (i.e., topology) when none
exists.})

The overall appearance of the halo is remarkably similar between the
codes, a major feature of the halo being its irregular shape. The left
side of the halo is elongated and a second major peak has developed on
the right, leading o a triangular shape in this projection. This
irregularity (seen also very clearly in Figure~\ref{halo3splash}) is
most likely the reason for the disagreement in the inner part of the
profiles. The halo has probably undergone a recent merger or is in the
process of merging. Comparing the lower resolution runs from MC$^2$
and FLASH with {\small GADGET-2} and HOT, the effect of force
resolution is very apparent, the high resolution runs producing
significantly more substructure. {\small GADGET-2} shows slightly more
substructure than HOT, which could be due to the adaptive time
stepping used in the {\small GADGET-2} run relative to HOT's global
time-step.

\begin{figure*}
\begin{center}
\parbox{12cm}{
\parbox[t]{12cm}{
\parbox[t]{6.cm}
{\begin{center}
\hspace{0cm}\includegraphics[width=60mm,angle=0]{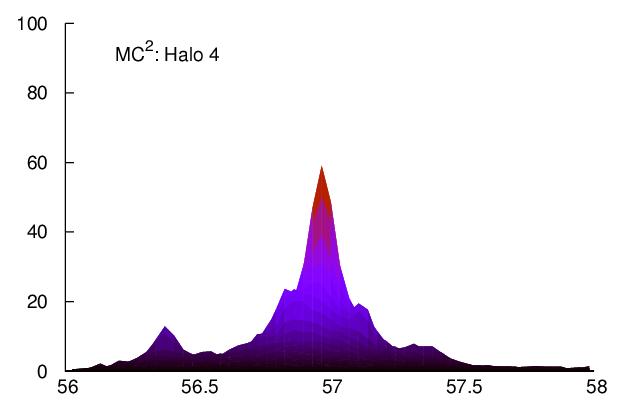}
\end{center}}
\parbox[t]{6.cm}
{\begin{center}
\hspace{0cm}\includegraphics[width=60mm,angle=0]{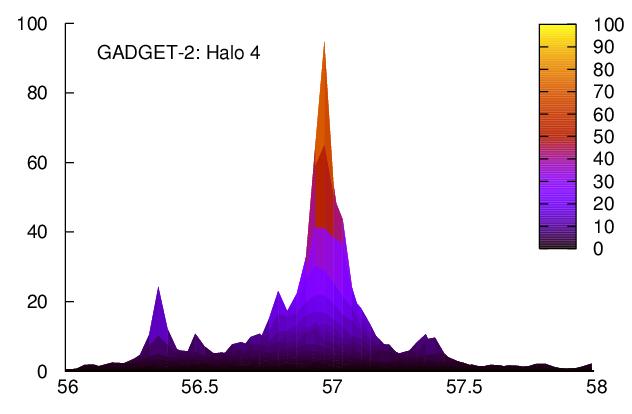}
\end{center}}}
\parbox[t]{12cm}{
\parbox[t]{6.cm}
{\begin{center}
\hspace{0cm}\includegraphics[width=60mm,angle=0]{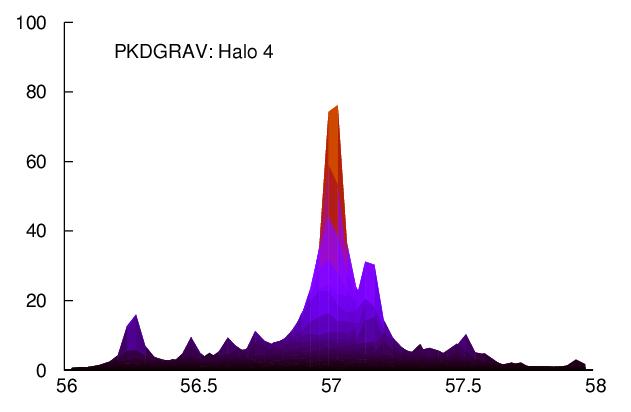}
\end{center}}
\parbox[t]{6.cm}
{\begin{center}
\hspace{0cm}\includegraphics[width=60mm,angle=0]{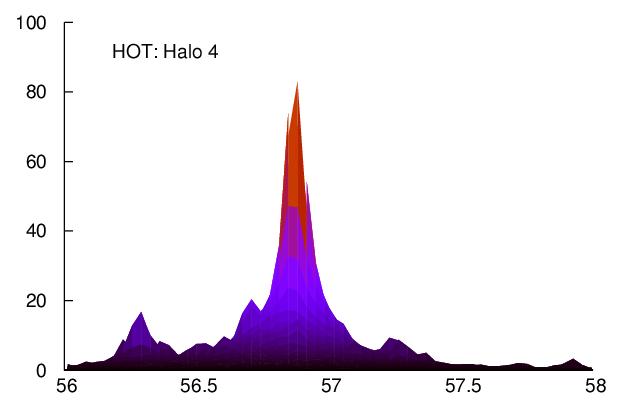}
\end{center}}}
}
\end{center}
\caption{\label{halo4}Two-dimensional density profile of Halo 4 for
MC$^2$, {\footnotesize GADGET-2}, PKDGRAV, and HOT. MC$^2$ shows less
substructure and is less dense in the inner region.}
\end{figure*}

Figure~\ref{halo3d} shows Halo 3 from the remaining six runs. As in
Figure~\ref{halo1}, the two-dimensional density is shown on a
100$\times$100 grid. The three-dimensional view underlines the rather
complicated structure of the halo. PMM and Enzo show the elongated
structure with two maxima, whereas the Hydra and PKDGRAV results
differ somewhat from the other codes. They have a more well
defined peak and do not exhibit much of the second structure. TreePM
and TPM are very similar to {\small GADGET-2} and HOT. Overall, Halo 3
has much more interesting features than Halo 1, which leads to slight
discrepancies in the halo profiles among the codes.

Last, we study Halo 4 from a subset of the codes: MC$^2$, {\small
  GADGET-2}, PKDGRAV, and HOT, covering the grid, tree-PM, and tree
codes. The results are shown in Figure~\ref{halo4}. As before, the
lower density of the PM code is due to its restricted
resolution. Overall, the agreement is again very satisfying. The
centers of the halos are in excellent agreement, and all four runs
show a smaller structure on the left of the main halo. The exact
details of the smallest structures are different which could be due to
inaccurate time-stepping and discrepancies in the codes' output
redshifts.

Overall, the comparison of the largest halos in the box is very
satisfactory. The halo profiles agree on the scales expected from the
code resolutions. Differences of the inner parts can be explained due
to very irregular shapes as in Halo 3. The reader should keep in mind
that we did not resimulate the halos with higher resolution, and that
these halos ere extracted straight out of a cosmological volume
simulation. Therefore, the level of agreement is in accord with
theoretical expectations.

\subsection{The Mass Function and Halo Counts as a Function of
Density}

\subsubsection{The Mass Function}

An important statistic in cosmology is the number count of halos as a
function of mass, the so-called mass function. The mass function of
clusters of galaxies from ongoing and upcoming surveys can provide
strong constraints on dark energy~\cite{majmohr}.  Numerous studies
have been carried out to predict the mass function
theoretically~\cite{ps,ls,stm}. Because halo formation is a strongly
nonlinear dynamical process, the approximations underlying analytic
predictions limit the attainable accuracy for constraining
cosmological parameters. Nevertheless, some of these analyses can help
to understand the origin and qualitative behavior of the mass
function.

In order to obtain more precise predictions, several groups have
carried out large N-body simulations to find accurate fits for the
mass function~\cite{st,jenkins,waht,reed03,reed07}.  In addition, the
evolution of the mass function has been studied in
detail~\cite{reed03,heitmann06,reed07,lukic07}. The numerical study of the
mass function poses several challenges to the simulation code,
especially if one wants to obtain reliable results at the few percent
accuracy level: the number of particles in a halo has to be sufficient
in order to prevent systematic biases in determinations of the halo
mass~\cite{waht}, the force resolution has to be adequate to capture
the halos of interest~\cite{heitmann06,lukic07}, the simulation has to
be started at sufficiently high redshift~\cite{heitmann06,lukic07},
and finite box corrections might have to be considered if the
simulation box is small~\cite{lukic07,bl,bp} (for a comprehensive
study of possible systematic errors in simulating the mass function
and its evolution, see~\cite{lukic07}).

\begin{figure}
\center\includegraphics[width=70mm]{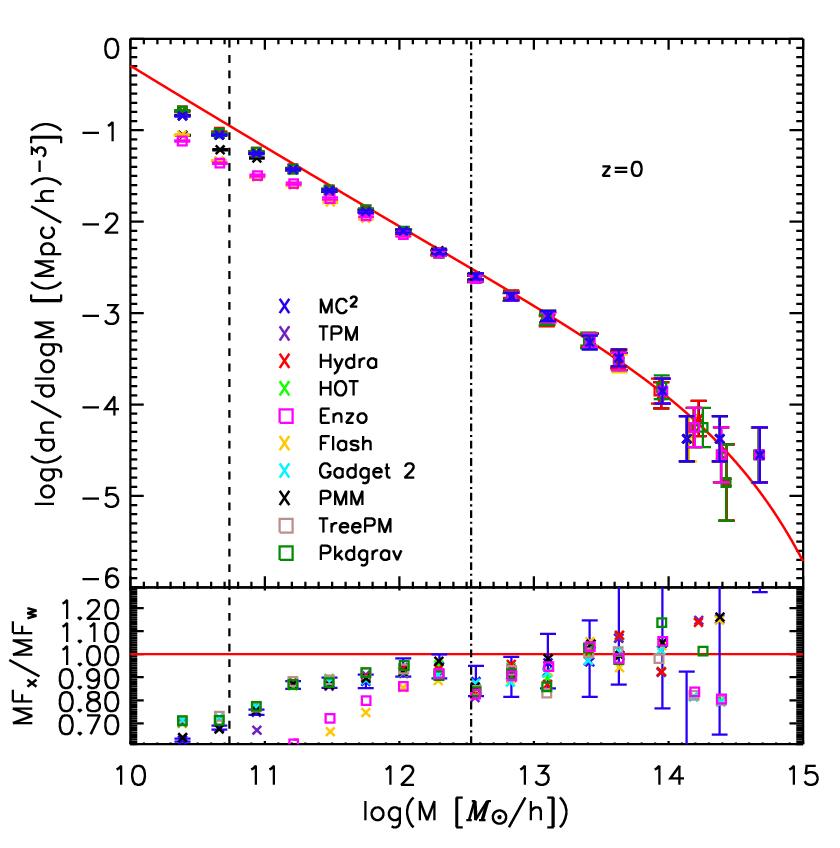}
\caption{\label{massf}Mass function at $z=0$, simulation 
results and the Warren fit (red line).  Lower panel: residuals with
respect to the Warren fit. For clarity we only show the error bars for
one code. The dashed line indicates the threshold for 40 particles
(force resolution limit for the PM codes, according to
Equation~(\ref{resol})), the dotted-dashed line for 2500 particles
(force resolution limit for the base grid of the AMR codes). }
\end{figure}

In this paper we study the mass function at $z=0$. We identify halos
with a friends-of-friends (FOF) algorithm~\cite{fof} with linking
length of $b=0.2$. The smallest halo we consider has 10 particles, not
because this is physically reasonable (usually the minimum number of
particles is several times bigger), but because we are interested in
cross-code comparison. We follow the suggestions by Warren et
al.~\cite{waht} and correct the halo mass for possible undersampling
via:
\begin{equation}
n_{h}^{\rm corr}=n_{h}\left(1-n_{h}^{-0.6}\right),
\end{equation}
where $n_h$ is the number of particles in a halo. This correction
lowers the masses of small mass halos considerably. 

In order for small halos to be resolved, both mass and force
resolution must be adequate. In Ref.~\cite{lukic07}, resolution
criteria for the force resolution are derived:
\begin{equation}
\label{resol}
\frac{\delta_f}{\Delta_p} < 0.62\left[\frac{n_h\Omega_m(z)}
{\Delta}\right]^{1/3},
\end{equation}
with $\delta_f$ being the force resolution, $\Delta_p$ being the
interparticle spacing, and $\Omega_m(z)$ the matter content of the
Universe at a given redshift. Equation~(\ref{resol}) predicts that all
the non-grid codes have enough force resolution to resolve the smallest
halos considered, while the two PM codes, MC$^2$ and PMM, have
sufficient force resolution to resolve halos with more than 40
particles, and that the base grid of the two AMR codes restricts them
to capturing halos with more than 2500 particles. Of course this is
only a rough estimate in principle since the AMR codes increase their
local resolution as a function of density threshold, the question is
whether the criteria used for this is sufficient to resolve halos
starting at 40 particles/halo.

We have indicated the resolution restrictions in Figure~\ref{massf} by
vertical lines (dashed: 40 particles, dashed-dotted: 2500
particles). The predictions are good indicators of actual code
results. The AMR codes fall off at slightly lower masses than given by
2500 particles. This shows that the resolution which determines the
smallest halos being captured is dictated by the base grid of the AMR
codes and not by the highest resolution achieved after
refinement. Thus, for the AMR codes to achieve good results,
significantly more aggressive density thresholding appears to be
indicated. (Similar results were found in
Refs.~\cite{Heitmann05,oshea05}.) As predicted, the mass functions of
the PM codes start to deviate at around 40 particles from the other
codes.

Overall the agreement among the codes is very good. For comparison, we
show the Warren fit~\cite{waht} in red. Due to limited statistics
imposed by the small box-size, the purpose here is not to check the
accuracy of the fitting function. At the high mass end, the scatter is
as expected due to the rareness of high-mass halos. In the medium mass
range between $10^{12.3}$ and $10^{13.4}h^{-1}$M$_\odot$ all codes
agree remarkably well, down to the percent level. In the small halo
regime with as low as 40 particles, the agreement of the codes --
besides the AMR codes as explained above -- stays at this level. This
indicates that the halo mass function is a very robust statistic and
the simple resolution arguments given above can reliably predict the
halo mass limits of the individual simulations.

The comparison yields one surprising result, however: the TPM code
simulation has far fewer halos in the regime below 40 particles per
halo than the other high resolution codes. This finding was already
pointed out in Ref.~\cite{Heitmann05}. In order to understand this
deficit of halos in more detail we investigate the halo count as a
function of environment in the following.

\subsubsection{Halo Count and Density}
In this section we use ParaView again as the main analysis tool.  One
very attractive feature of ParaView is a suite of filter functions.
These filters allow direct manipulation of the data that is
visualized. They include functions such as Fast Fourier Transforms,
smoothing routines via Gaussian filtering (which we used in the
previous section), and tesselation routines, to name a few. We have
implemented additional routines to find halos (a fully parallel FOF
halo finder integrated into ParaView is under development) and to
calculate densities from the particle distribution in order to
cross-correlate density with halo counts. We have also added an
interface to the plotting program gnuplot.\footnote{These new routines
  are not yet available in the public version of ParaView but we plan
  to release them in the near future.}

\begin{figure*}
\begin{center}
\parbox{15cm}{
\parbox[t]{15cm}{
\parbox[t]{4.3cm}
{\begin{center}
\hspace{0cm}\includegraphics[width=43mm]{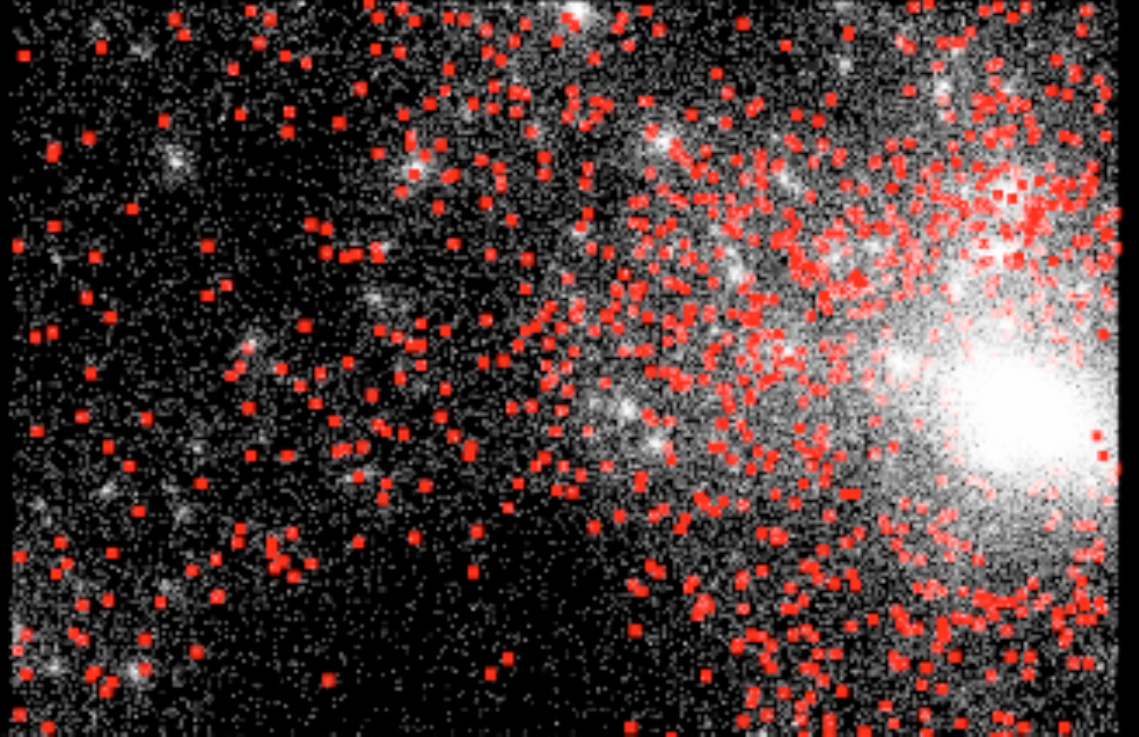}
\end{center}}
\parbox[t]{4.3cm}
{\begin{center}
\hspace{0cm}\includegraphics[width=43mm]{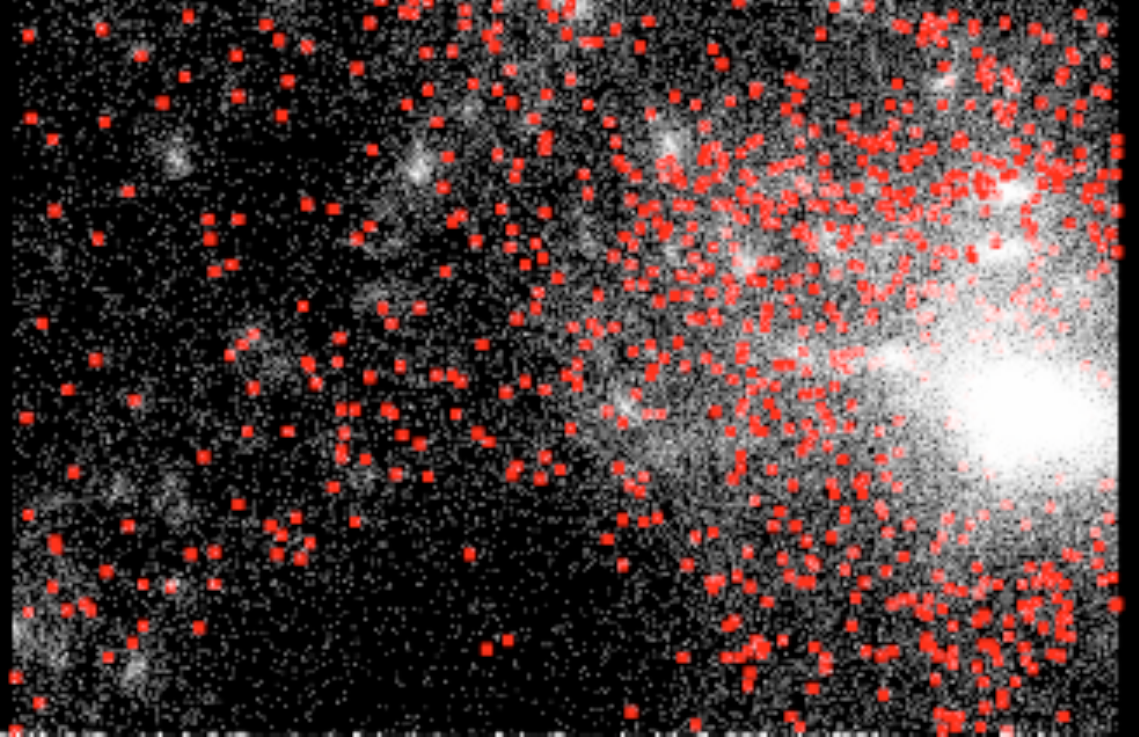}
\end{center}}
\parbox[t]{4.3cm}
{\begin{center}
\hspace{0cm}\includegraphics[width=43mm]{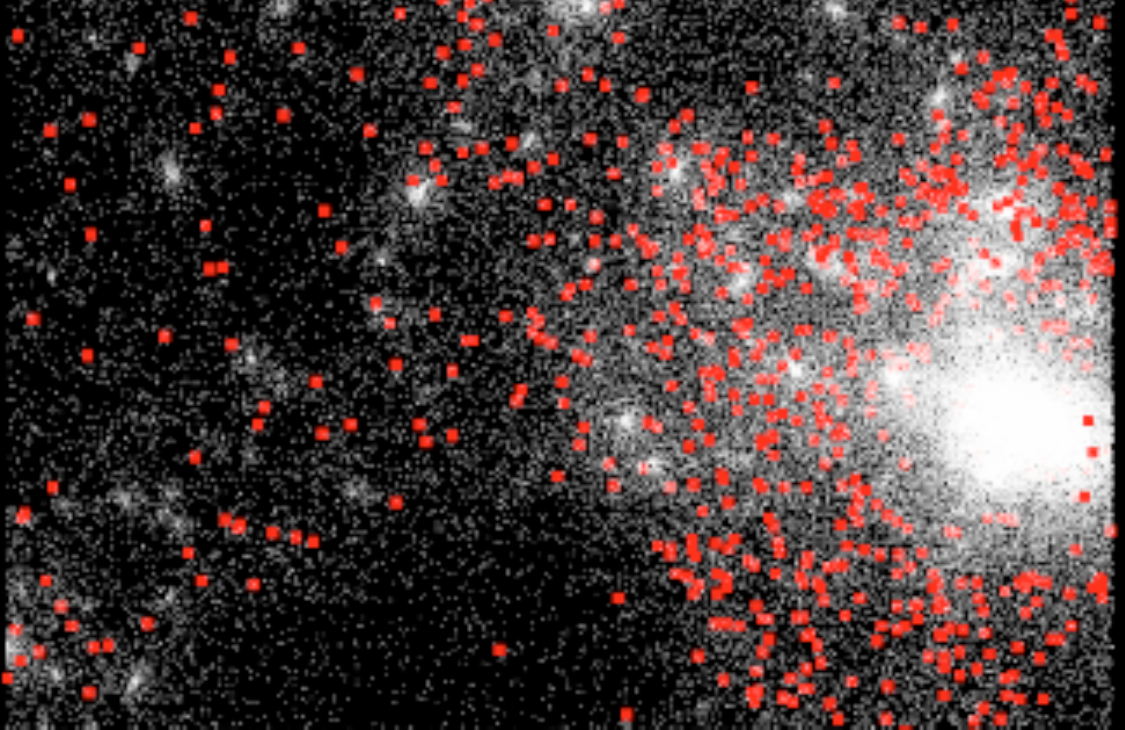}
\end{center}}
}}
\end{center}
\caption{Small halos (10 particles) in the HOT, MC$^2$, and TPM 
simulation. Red points: halos, white dots: subset of the simulation
particles. The distribution and number count of the small halos is
different in all three codes.}
\label{fig:smallhalos}
\end{figure*}

In the last section we investigated the mass function and discovered a
discrepancy of small halos in the two AMR codes and TPM. The
hypothesis for the halo deficit in the AMR codes is, as discussed
above, that the base grid resolution is too low and allows us only to
catch halos with more than 2500 particles accurately. The coarse base
grid in the initial state of the simulation does not allow for small
halos to form and these halos cannot be recovered in the end. This
would imply that the AMR simulations should have a deficit of small
halos more or less independent of density: small halos should be
missing everywhere, even in the highest refinement regions. A possible
explanation for the missing halos in the TPM simulation could be a
hand-over problem between the PM and the tree code. In this case, the
number of small halos in high density regions should be correct. A
qualitative comparison of three codes (HOT, MC$^2$, and TPM) is shown
in Figure~\ref{fig:smallhalos}.  The red points show halos with 10
particles, the white dots are a subset of the simulation particles. It
is immediately obvious, that the halo counts in different
environments, close to the large halo on the right, or on in the lower
density regions on the left, are different. After this qualitative
result, we have to quantify this finding in order to come to a
reliable conclusion about the cause for the halo deficits.

\begin{figure*}
\begin{center}
\parbox{14cm}{
\parbox[t]{14cm}{
\parbox[t]{7cm}
{\begin{center}
\hspace{0cm}\includegraphics[width=70mm,angle=0]{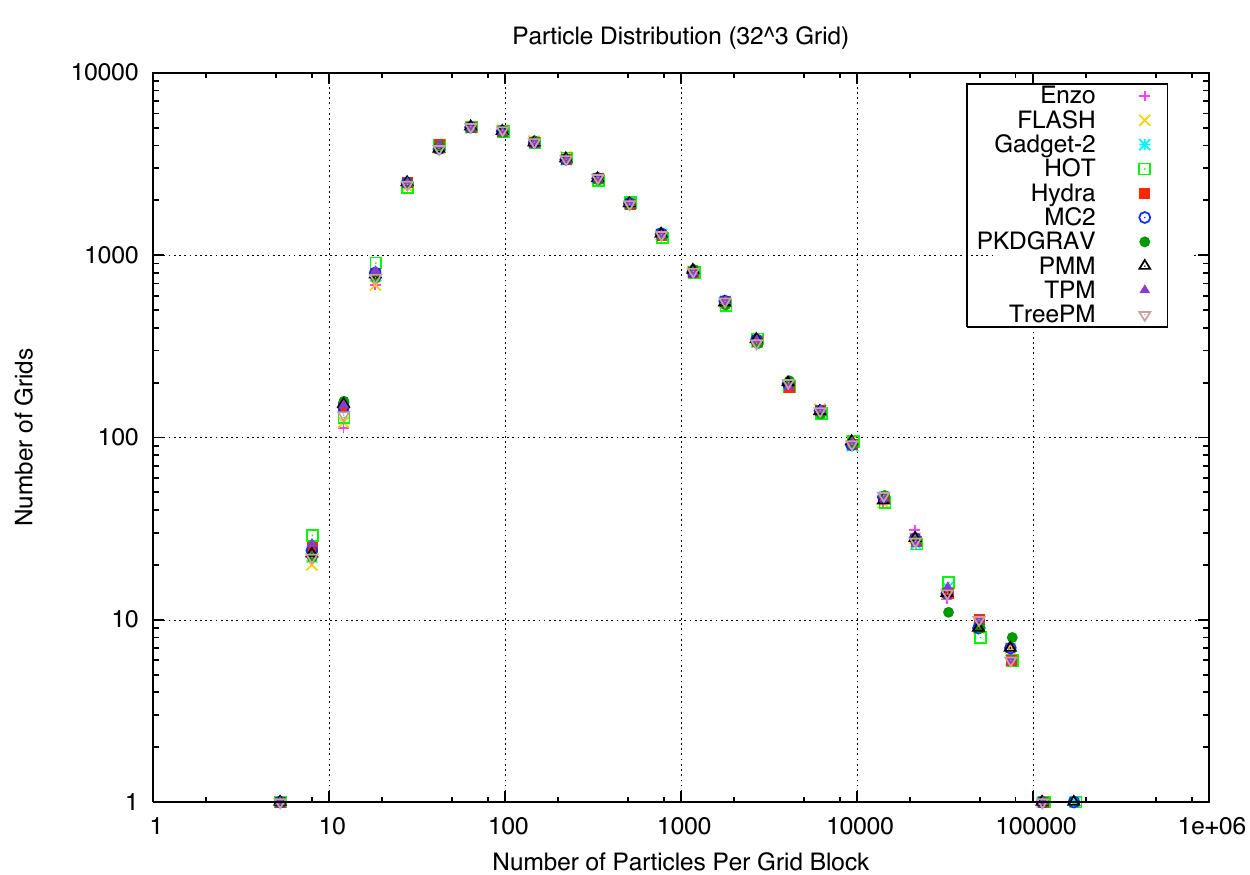}
\end{center}}
\parbox[t]{7cm}
{\begin{center}
\hspace{0cm}\includegraphics[width=70mm,angle=0]{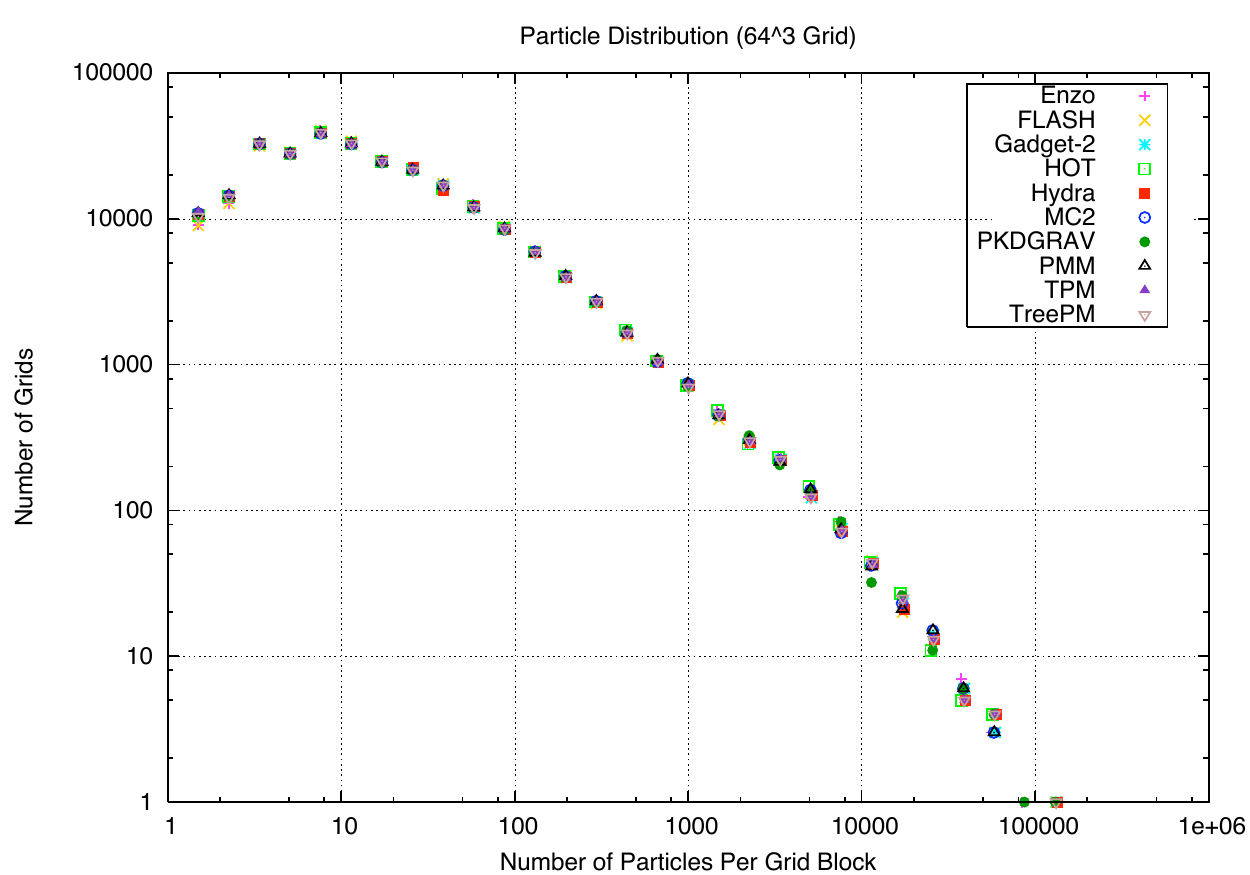}
\end{center}}}
\caption{\label{fig:pdf}Probablity distribution function of the
densities. Left panel: calculation of the density on a $32^3$ grid,
right panel: calculation of the density on a $64^3$ grid.}}
\end{center}
\end{figure*}

We use the VTK toolkit to implement a routine that calculates the
density field on a (variable) grid from the particle distribution via
a nearest grid point (NGP) algorithm. The grid size for the density
field is usually set by the requirement that the density field be not
too noisy. As a first check we compare the density probability
distribution function (PDF) for the different codes. It is clear that,
if the grid for calculating the density is chosen coarse enough,
details should be smoothed out and the PDFs for the different codes
should be in good agreement. In Figure~\ref{fig:pdf} we show the PDFs
for all codes calculated on a $32^3$ grid (left panel) corresponding
to a smoothing scale of 2$h^{-1}$Mpc and a $64^3$ grid (right panel)
corresponding to a smoothing scale of 1$h^{-1}$Mpc. In both cases all
codes agree extremely well, as to be expected since the smoothing
scales are well beyond the code resolutions. We confirmed that this
result holds also for finer grids, up to 256$^3$, which corresponds to
the lowest resolution in the AMR codes Enzo and FLASH. The average
number of particles in a grid cell $\bar\rho$ on the left panel is 512
particles per cell, in the right panel 64 particles per cell. If we
define the density contrast $\delta=(\rho-\bar\rho)/\bar\rho$ and
define void (highly underdense) regions as regions with a density
contrast $\delta^{\rm Void}=-0.8$, we find $\rho^{\rm Void}\simeq 100$
for the left panel and $\rho^{\rm Void}\simeq 13$ for the right
panel. In both cases, this threshold is on the right of the maximum of
the curves -- a large fraction of the simulation volume is underdense.

\begin{figure*}
\center\includegraphics[width=150mm]{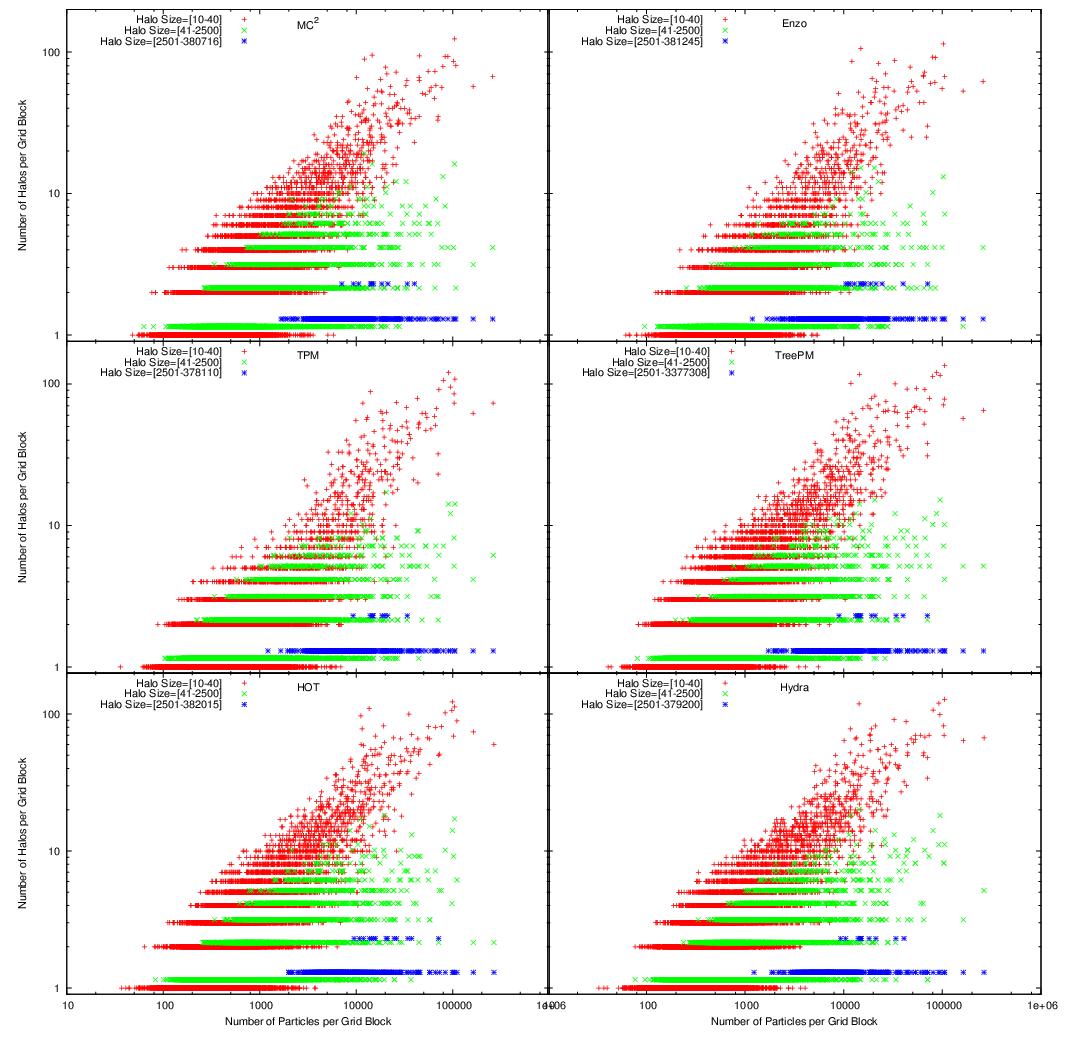}
\caption{\label{corr}Correlation between the particle density
(calculated on a $32^3$ grid) and the numbers of halos in a certain
density regime. The smallest halos (10-40 particles) are marked in
red, medium size halos (41-2500 particles) are shown in green, and the
biggest halos (more than 2500 particles) are shown in blue. The green
crosses are shifted by 0.15 and the blue crosses by 0.3 upwards to
make all points visible.}
\end{figure*}

Next we investigate the correlation of the numbers of halos with
density (measured by the number of particles per grid cell on a $32^3$
grid). The results for six codes are shown in Figure~\ref{corr}. The
results for the four other codes are very similar compared to the
codes using the same algorithm (e.g. the FLASH result is very close to
the Enzo result). The scatter plots show three different mass bins:
very light halos (10-40 particles) in red, medium mass halos (41-2500
particles) in green, and the heaviest halos in the simulations (2500
and more particles) in blue.  These thresholds were chosen because, as
discussed earlier, the force resolution of MC$^2$ and PMM should be
sufficient to resolve halos with more than 40 particles, while Enzo's
and FLASH's base grid set this limit to more than 2500 particles. The
green and blue crosses are slightly shifted upward to make all points
visible. The cut-offs on the left of the scatter plots are easy to
understand: a certain minimum density is required to have a halo of a
certain size and a certain number of halos, e.g. 50 small halos cannot
live in an underdense cell. Enzo and TPM show the largest deficit of
halos, confirming the mass function results. The small halos in Enzo
are mainly missing in the low density regions, and below $\delta^{\rm
Void}$ there are almost no halos. TPM has a very large deficit between
1,000-10,000 particles per cell, corresponding to a density contrast
$\delta$ between 1 and 20. This confirms the visual impression from
Figure~\ref{fig:smallhalos}. The overall agreement especially for the
larger halos is very good. The largest halo (blue cross on the far
right) seems to be surrounded by a large number of small halos,
consistent among all codes.

\begin{figure*}
\begin{center}
\parbox{15cm}{
\parbox[t]{15cm}{
\parbox[t]{7.5cm}
{\begin{center}
\hspace{0cm}\includegraphics[width=75mm,angle=0]{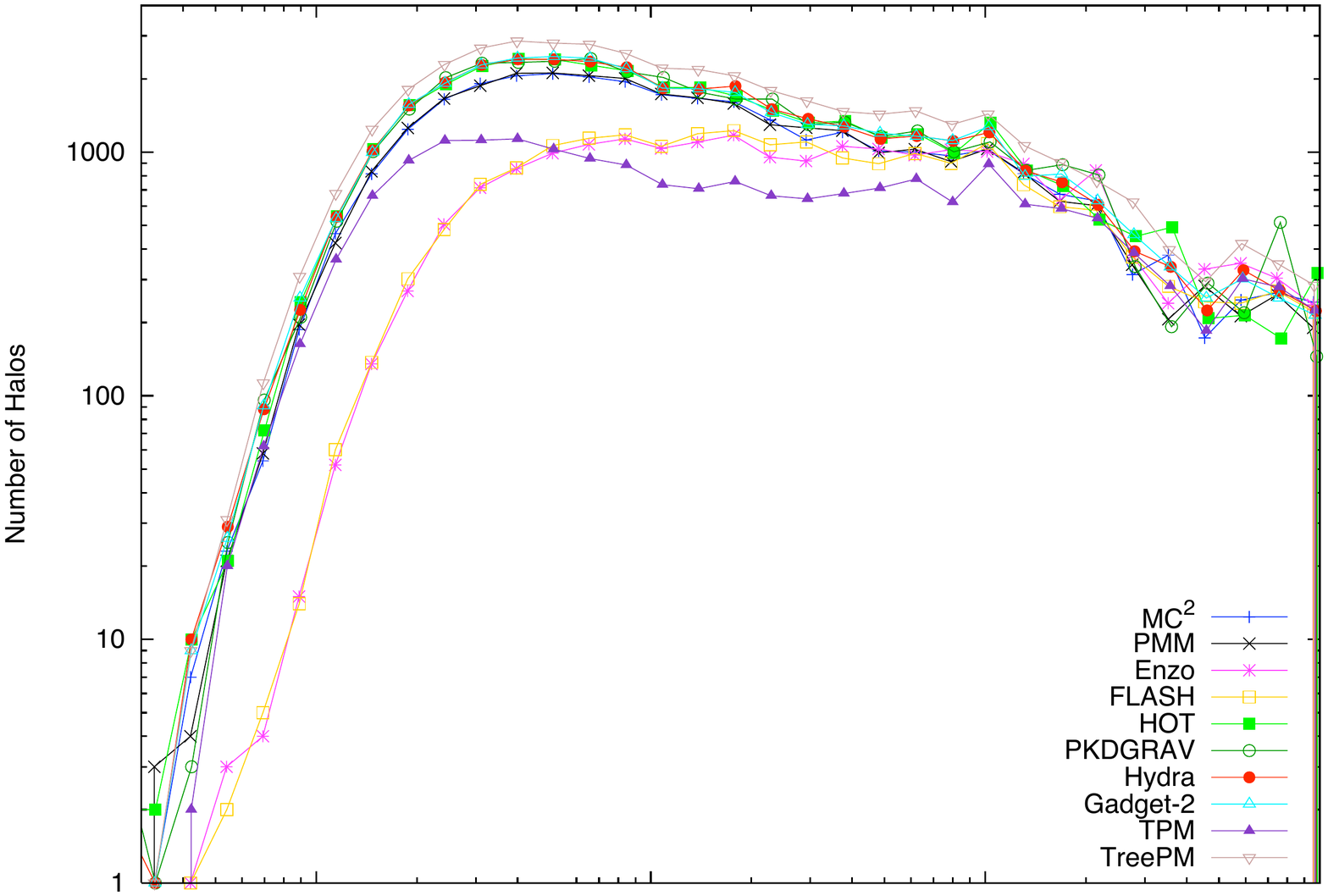}
\end{center}}
\parbox[t]{7.5cm}
{\begin{center}
\hspace{0cm}\includegraphics[width=75mm,angle=0]{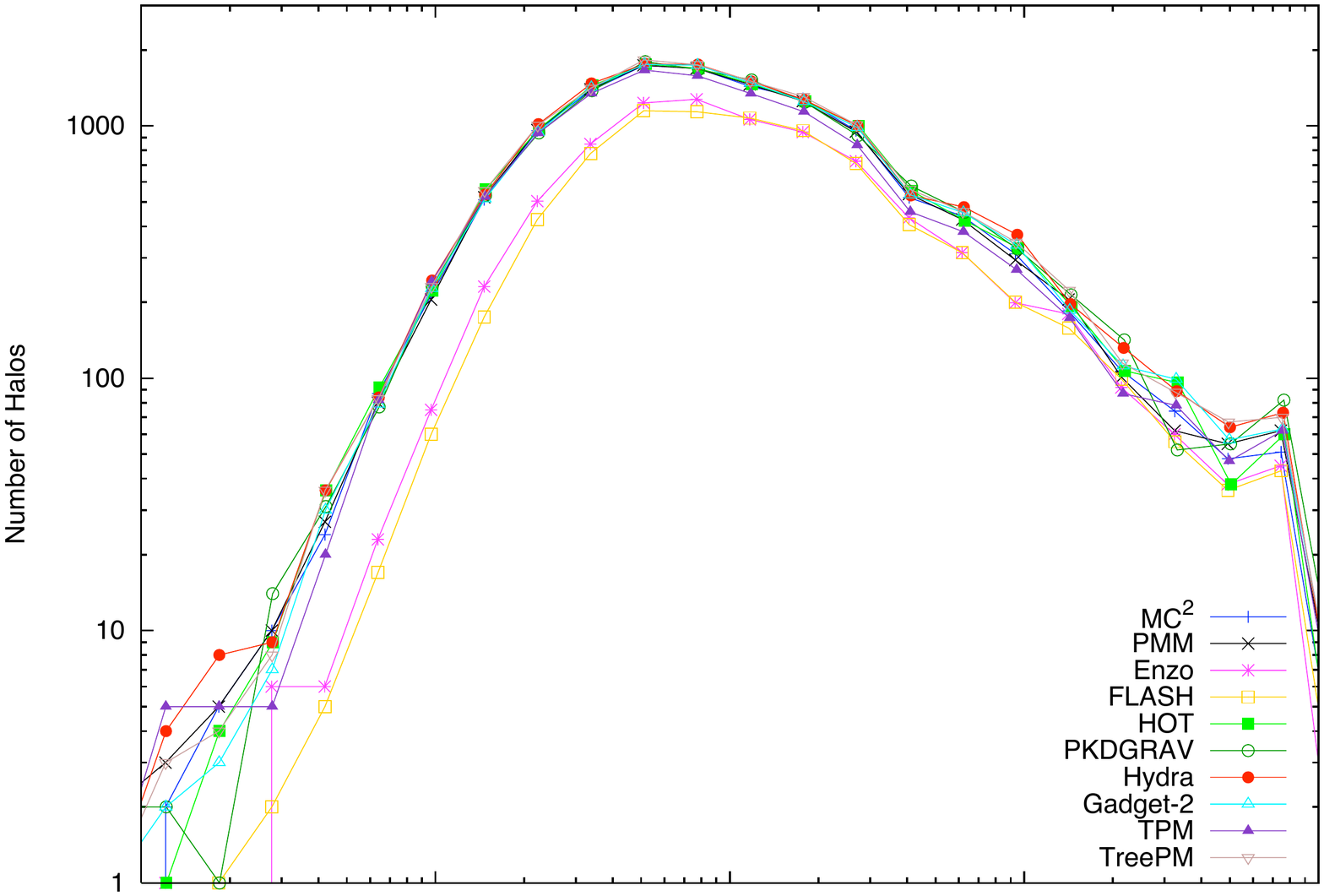}
\end{center}}}

\vspace{-1.9cm}

\parbox[t]{15cm}{
\parbox[t]{7.5cm}
{\begin{center}
\hspace{0cm}\includegraphics[width=75mm,angle=0]{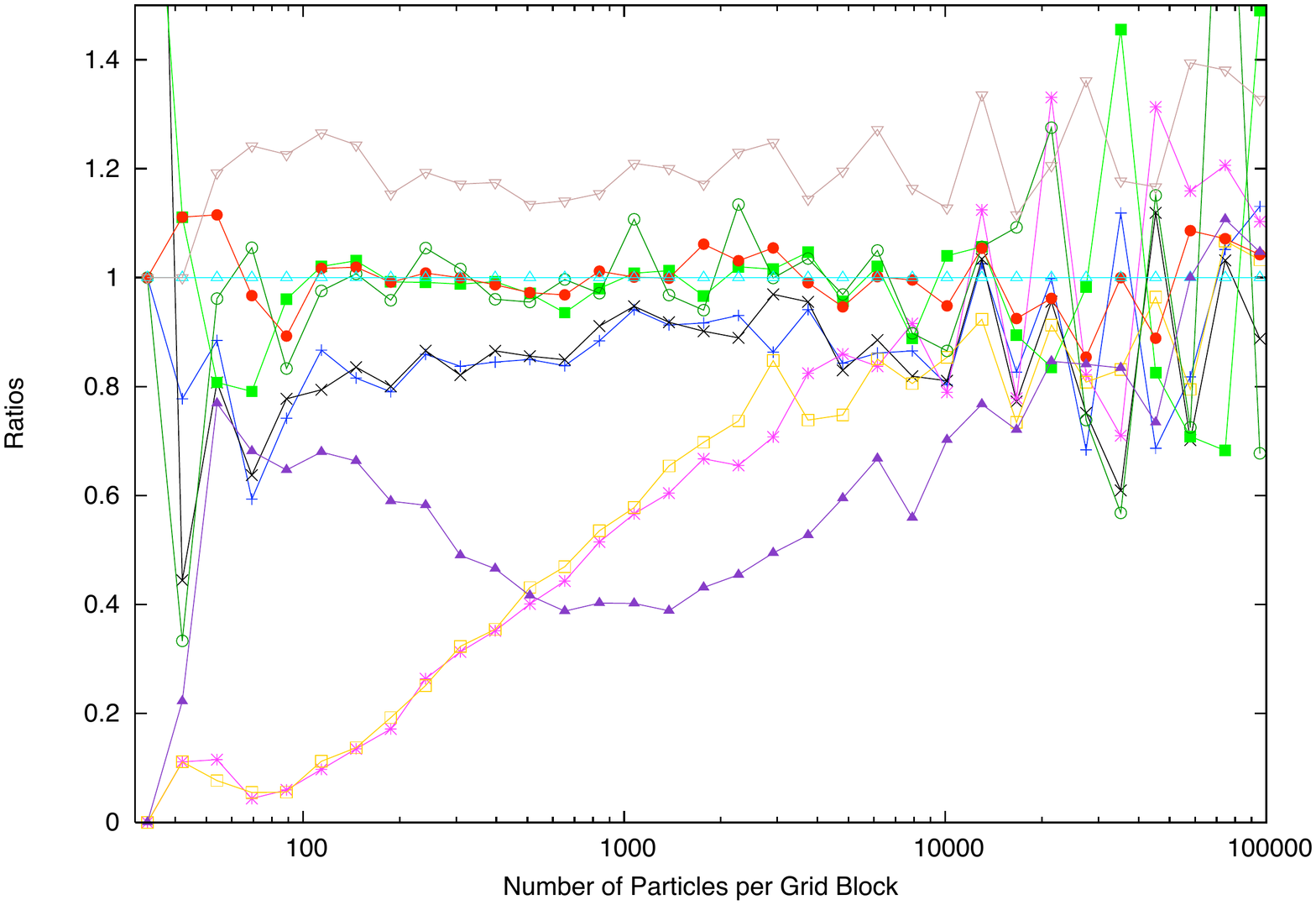}
\end{center}}
\parbox[t]{7.5cm}
{\begin{center}
\hspace{0cm}\includegraphics[width=75mm,angle=0]{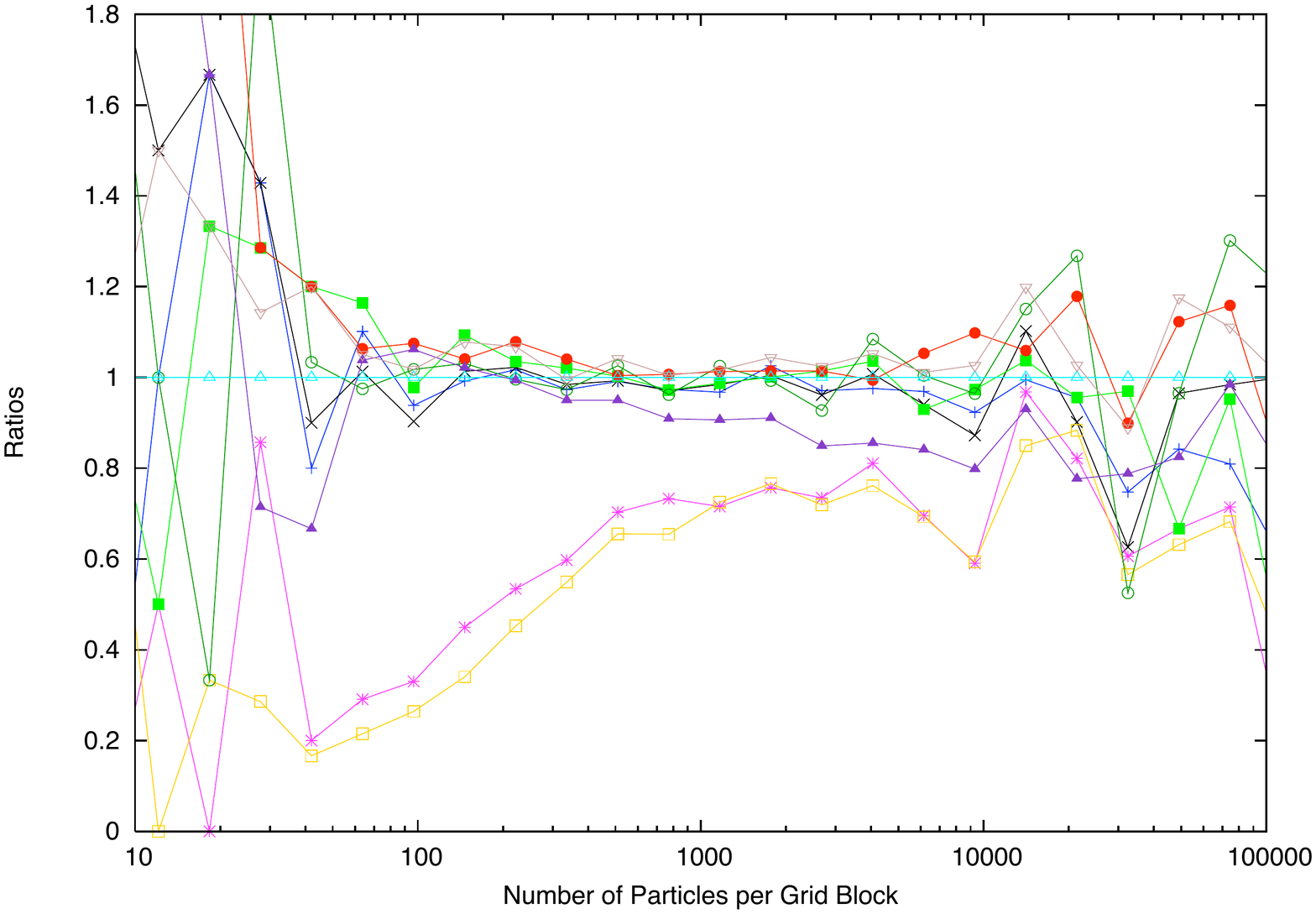}
\end{center}}}
}
\end{center}
\caption{\label{fig:histo} Number of halos as a function of
  density. Left panel: halos with 10 - 40 particles, right panel:
  halos with 41 - 2500 particles. The lower panels show the residuals
  with respect to {\footnotesize GADGET-2}. Both panels show the
  deficit of small halos in Enzo and FLASH over most of the density
  region -- only at very high densities do the results catch up. The
  behavior of the TPM simulation is interesting: not only does this
  simulation have a deficit of small halos but the deficit is very
  significant in medium density regions, in fact falling below the two
  AMR codes. The slight excess of small halos shown in the TreePM run
  vanishes completely if the halo cut is raised to 20 particles per
  halo and the TreePM results are in that case in excellent agreement
  with {\footnotesize GADGET-2}. }
\end{figure*}

To cast the results from Figure~\ref{corr} in a more quantitative
light, Figure~\ref{fig:histo} displays the distribution of halos with
respect to density for the two lower mass bins. From
Figure~\ref{corr} we can read off the density range of interest for
each mass bin, i.e. the density range with the largest halo
population. We restrict our investigations to a density threshold of
up to 100,000 particles per cell. Figure~\ref{fig:histo} shows the
results for 10-40 particle (left panel) and 41-2500 particle halos
(right panel). The lower panels show the residuals with respect to
{\small GADGET-2}. (We have verified that the agreement for larger
halos between the ten codes is very good as expected from
Figure~\ref{corr}.) The two AMR codes Enzo and FLASH have a deficit
for both halo sizes over most of the density region. They only catch
up with the other codes at around 10,000 particles per cell, in
agreement with the our previous argument that whether halos are
resolvable by the AMR codes or not is dictated by the size of the base
grid. In terms of capturing smaller halos, the refinement only helps
in very high density regions.

The result for the TPM simulation is somewhat paradoxical: in the low
density region the result for the small halos agrees well with the
other high-resolution codes, however, TPM misses a very large number
of small halos in the region between 200 and 10,000 particles per
cell, the curve falling even below the AMR codes. This suggests that
the problem of the TPM code is not due to the threshold criterion for
the tree but perhaps due to a hand-over problem between the grid and
the tree. The two PM codes have slightly lower numbers of very small
halos, in good agreement with the prediction that they only resolve
halos with more than 40 particles. The agreement between MC$^2$ and
PMM itself is excellent. The TreePM code shows a slight excess of
small halos compared to the other high-resolution codes. This excess
vanishes completely if the cut for the small halos is chosen to be 20
particles instead of 10 particles for the smallest allowed halo. This
indicates a slightly higher force resolution in the TreePM run
compared to the other runs. The agreement for the medium size halos
(left panel) is very good, except for the AMR codes. For the medium
size halos, the TPM code again shows a slight deficit of halos in the
medium density regime, but far less pronounced than for the small
halos. The overall agreement of the high-resolution codes is very
good, as is to be expected from the mass function results.

\subsection{The Power Spectrum}

The matter power spectrum is one of the most important statistics for
precision cosmology. Upcoming weak lensing surveys promise
measurements of the power spectrum at the one percent accuracy level
out to length scales of $k\sim 10 h$Mpc$^{-1}$ (for an overview of the
requirements for the accuracy of predictions for future lensing
surveys, see, e.g., Ref.~\cite{weaklensing}). This poses a severe
theoretical challenge: predicting the matter power spectrum at the
same level of accuracy. A first step for showing that this is possible
is to investigate how well the matter power spectrum can be predicted
from pure dark matter simulations, baryonic physics being included as
a second step. It has already been shown that at the length scales of
interest, hydrodynamic effects can alter the matter power spectrum at
up to 10 percent~\cite{baryon}. In this paper we concentrate on the
first step and determine how well a diverse set of N-body codes agree
with each other for the prediction of the matter power spectrum. In
future work we aim to predict the dark matter power spectrum at $k\sim
1 h$Mpc$^{-1}$ at the level of one percent accuracy or better. This
will include a detailed analysis of the accuracy of the initial
conditions as well as of the nonlinear evolution, a task beyond the
scope of the current paper.

\begin{figure}[t]
\center\includegraphics[width=120mm]{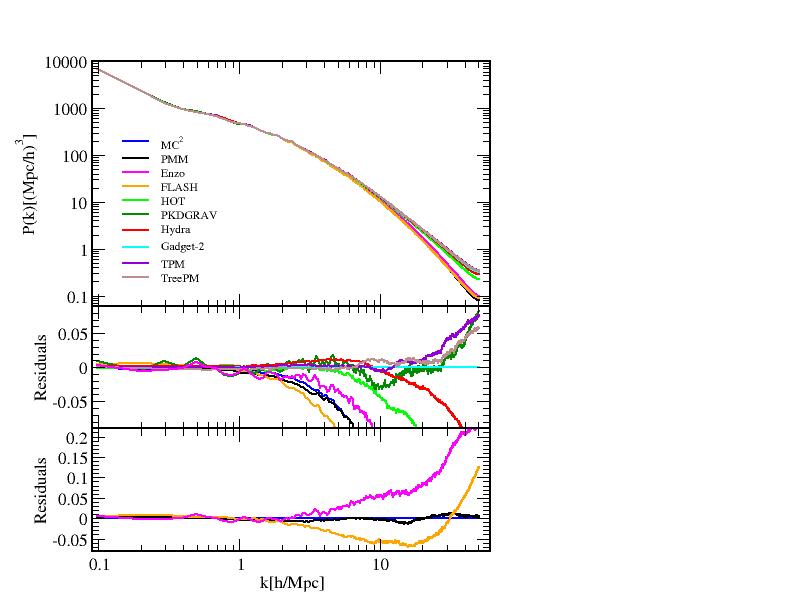}
\caption{\label{pow}Power spectrum results and the residuals for the different
codes.  Upper panel: comparison of the different power spectra. Middle
panel: residuals of all codes with respect to {\footnotesize GADGET-2}. Lower
panel: Residuals of the mesh codes with respect to MC$^2$.}
\end{figure}

We determine the matter power spectrum by generating the density field
from the particles via a Cloud-in-Cell (CIC) routine on a 1024$^3$
spatial grid and then obtain the density in $k$-space by applying a
1024$^3$ FFT. The square of the $k$-space density yields the power
spectrum: $P(k)=\langle|\delta(k)^2|\rangle$.  The CIC routine
introduces a filter at small length scale. We compensate for this
filtering artifact by deconvolving the $k$-space density with a CIC
window function.

The results for the different codes are shown in
Figure~\ref{pow}. Note that the box size of $64 h^{-1}$Mpc is too
small for a realistic cosmological power spectrum calculation, as the
largest modes in the box no longer evolve linearly. This leads to an
incorrect onset of the nonlinear turn-over in the power
spectrum. Nevertheless, the comparison of the different codes is very
informative. The upper panel in Figure~\ref{pow} shows the results for
the power spectra themselves. The lower resolution of the grid codes
is very apparent, their results falling away at $k\sim 2
h$Mpc$^{-1}$. The middle panel shows the residuals of all codes with
respect to {\small GADGET-2}. All codes agree at roughly 1\% out to
$k\sim 1 h$Mpc$^{-1}$. PKDGRAV shows small scatter in the linear
regime.  This might be caused by imprecise periodic boundary
conditions, which are not as easy to implement in tree codes as they
are for grid codes.  The high-resolution codes agree to better than
5\% out to $k\sim 10 h$Mpc$^{-1}$. At that point HOT and Hydra lose
power, while PKDGRAV, TPM, and TreePM show slightly enhanced power
compared to the {\small GADGET-2} run. The formal force resolutions of
the codes would suggest that the different runs (including the grid
runs) should agree much better at the wavenumbers shown.

The 1024$^3$ FFT used to generate the power spectra is far below the
resolution of the non-grid codes and at the resolution limit of the
AMR and PM codes. The discrepancy might be due to several reasons: the
number of time steps, the accuracy of the force solvers, the accuracy
of reaching $z=0$ at the end of each run, just to suggest a few. A
more detailed study of the power spectrum including larger simulation
boxes is certainly required to obtain the desired accuracy for
upcoming surveys.  In the lower panel we show a comparison of the grid
codes only, with respect to MC$^2$. The two pure PM codes, MC$^2$ and
PMM agree remarkably well over the whole $k$-range under
consideration, the difference being below 1\%. The two AMR codes,
Flash and Enzo, deviate considerably, most likely due to different
refinement criteria. It is somewhat surprising that Enzo has larger
power than the two PM codes, which have the same resolution in the
whole box that Enzo has only in high density regions. This could be
the result of an algorithmic artifact in the AMR implementation.

To summarize, the agreement for the matter power spectrum is at the
5\% level over a large range in length scale. The early deviation of
the grid codes is surprising, as the nominal resolution of all codes
should have been sufficient to generate agreement over a wider
$k$-range. In order to be able to obtain more cosmologically relevant
results at $k\sim 1h$Mpc$^{-1}$, much larger simulation boxes have to
be compared. At higher wavenumbers, baryonic effects become important
leading to the necessity of a much more involved comparison set-up.

\section{Conclusion and Discussion}
\label{conclusions}

The new era of precision cosmology requires new standards for the
reach and accuracy of large cosmological simulations. While
previously, qualitative answers and quantitative results at the 20\%
accuracy level were sufficient, we now need to robustly predict
nonlinear physics at the 1\% accuracy level. This demanding task can
only be achieved by rigorous code verification and error control.

In this paper we have carried out a comprehensive code comparison
project with 10 state-of-the-art cosmological simulation
codes. ParaView was introduced as a powerful analysis tool which
should make it more convenient for code developers to compare
results. In particular, results from the current suite of simulations
can function as a good database for reference purposes and
benchmarking.  The initial conditions are publicly
available\footnote{http://t8web.lanl.gov/people/heitmann/arxiv}.

The results from the code comparisons are satisfactory and not
unexpected, but also show that much more work is needed in order to
attain the required accuracy for upcoming surveys. The halo mass
function is a very stable statistic, the agreement over wide ranges of
mass being better than 5\%. Additionally, the low mass cutoff for
individual codes can be reliably predicted by a simple criterion.

The internal structure of halos in the outer regions of $\sim R_{200}$
also appears to be very similar between different simulation
codes. Larger differences between the codes in the inner region of the
halos occur if the halo is not in a relaxed state: in this case, time
stepping issues might also play an important role (e.g. particle orbit
phase errors, global time mismatches). For halos with a clear single
center, the agreement is very good and predictions for the fall-off of
the profiles from resolution criteria hold as expected. The
investigation of the halo counts as a function of density revealed an
interesting problem with the TPM code, the simulation suffering from a
large deficit in medium density regimes. The AMR codes showed a large
deficit of small halos over almost the entire density regime, as the
base grid of the AMR simulation set too low a resolution limit for the
halos.

The power spectrum measurements revealed definitely more scatter among
the different codes than expected. The agreement in the nonlinear
regime is at the 5-10\% level, even on moderate spatial scales around
$k=10 h$Mpc$^{-1}$. This disagreement on small scales is connected to
differences of the codes in the inner regions of the halos.

\section*{Acknowledgments} 

The calculations described herein were performed in part using the
computational resources of Los Alamos National Laboratory.  A special
acknowledgment is due to supercomputing time awarded to us under the
LANL Institutional Computing Initiative.

\end{document}